\documentclass[pra,twocolumn,showpacs,preprintnumbers,amsmath,amssymb,superscriptaddress]{revtex4}

\usepackage{graphicx,amsmath,amssymb,amsfonts,latexsym,color,dcolumn,bm,epsfig,subfigure,verbatim}
\usepackage{epsfig}
\usepackage{bm}
\usepackage{dsfont}

\def\bbm[#1]{\mbox{\boldmath $#1$}}

\newcommand{\KK}{{\mathbf{K}}}
\newcommand{\qq}{{\mathbf{q}}}
\newcommand{\rr}{{\mathbf{r}}}
\newcommand{\xx}{{\mathbf{x}}}
\newcommand{\eee}{{\mathbf{e}}}
\newcommand{\uu}{{\mathbf{u}}}
\newcommand{\RR}{{\mathbf{R}}}
\newcommand{\kk}{{\mathbf{k}}}

\newcommand{\be}{\begin{equation}}
\newcommand{\ee}{\end{equation}}
\newcommand{\bea}{\begin{eqnarray}}
\newcommand{\eea}{\end{eqnarray}}

\usepackage{color}
\usepackage[english]{babel}

\def\bbm[#1]{\mbox{\boldmath $#1$}}

\begin{document}
\title{Photonic band-gap in a realistic atomic diamond lattice: penetration depth, finite-size and vacancy effects}

\author{Mauro Antezza}
\affiliation{Universit\'{e} Montpellier 2, Laboratoire Charles Coulomb UMR 5221 - F-34095, Montpellier, France}
\affiliation{CNRS, Laboratoire Charles Coulomb UMR 5221 - F-34095, Montpellier, France}

\author{Yvan Castin}
\affiliation{Laboratoire Kastler Brossel, ENS, UPMC and CNRS, 24 rue Lhomond 75231 Paris Cedex 05, France}

\date{\today}

\begin{abstract}
We study the effects of finite size and of vacancies on the photonic band gap recently predicted for an atomic diamond lattice. Close to a $J_g=0\to J_e=1$ atomic transition,  and for atomic lattices containing up to $N\approx 3\times10^4$  atoms, we show how the density of states can be affected by both the shape of the system and the possible presence of a fraction of unoccupied lattice sites. We numerically predict and theoretically explain the presence of shape-induced border states and of vacancy-induced localized states appearing in the gap. 
We also investigate the penetration depth of the electromagnetic field which we compare to the case of an infinite system. 
\end{abstract}


\maketitle

\section{Introduction}

That of waves propagation in periodic potentials constitute a problem shared by several domains of classical and quantum physics, ranging from the study of electron motion in metals \cite{Pastori}, to that of $X$- and $\gamma$-ray scattering by crystals \cite{Hopfield58,Pastori,Kagan}, and of light by photonics crystals and metamaterials \cite{bookDidier}. 
Periodicity leads to an organization of modes according to bands, and to the possible presence of band gaps, i.e.\ energy intervals where modes are absent. A periodic system is by definition infinitely extended, hence not physical. 
Nonetheless, predictions made on the base of infinite systems can become really satisfactory for systems large  enough, as in solid-state physics, and present the advantage to benefit from the 
Bloch theorem, and to be solved in the reciprocal space avoiding typical real space oscillating functions. 
Models based on infinite systems may however present some subtleties related to the way in which the infinite limiting process is 
performed, often requiring {\sl ad hoc} Ewald's summations type strategies. 

The recent experimental realization of a Mott phase with ultracold atomic gases \cite{Bloch02,Phillips07}, i.e.\ of artificial crystals made by single atoms trapped at the nodes of laser optical lattices, leads to the necessity of understanding the features of the band structure of light interacting with such systems. The peculiarity of this new system is that it presents several remarkable features: incident light scatters on point-like elementary quantum objects with an internal energy level structure, and a quantum delocalized position in space \cite{ACPRL2009}; the lattice periodicity is of the order of the incident light wavelength, allowing the exploration of the entire Brillouin zone and hence of possible band gaps \cite{ACPRA2009}; experiments reached a remarkable accuracy and control, permitting the realization of ultra-precise atomic clocks \cite{ultraclock1,ultraclock2,ultraclock3}. First attempts toward the description of such a system overlooked divergence problems, resulting in non correct prediction of band gaps \cite{Coevorden96,LagendijkRMP}, or were based on a {\sl ad hoc} ultraviolet regularisation 
procedure allowing to explore only a particular class of lattice geometries not presenting any band gap \cite{Knoester06}. 
Photonic band gaps of $1\mathrm{D}$ cold atomic vapors have been realized \cite{1DGuerin}, and exploited to generate optical parametric oscillation with distributed feedback \cite{1DGuerinlaser}. Scattered photons have been suggested  as a signature of the Mott insulator and superfluid quantum states \cite{Morigi10}, and studied in the framework of polaritons \cite{Carusotto08}, excitons and cavity polaritons \cite{Ritsch07} . Recently, by exploiting a microscopic theory of light-atom interaction \cite{Morice95}, and by explicitly introducing the presence of the unavoidable atomic quantum motion, it was possible to naturally regularize the divergences in a way independent of the lattice geometry, and at the same time to study the quantitative effects of the quantum atomic motion on the band structure \cite{ACPRL2009}. The explicit dependence of the photonic band structure on quantum features, as the atomic internal energy levels and the external atomic quantum motion, allows to consider this artificial structure as an example of quantum metamaterial \cite{SPIE}.  In the framework of the Fano-Hopfield self-consistent quadratic theory \cite{Hopfield58,Fano56,Carusotto08}, it was also possible to find an exact solution valid for the full Brillouin zone and for arbitrary Bravais and non-Bravais lattices, allowing the prediction of the diamond as the first  $3\mathrm{D}$ atomic lattice geometry presenting a complete photonic band  gap \cite{ACPRA2009}
\footnote{Although an optical diamond lattice was to our knowledge not realized
yet in the lab, the technique to be applied, elaborating on the ideas of \cite{Courtois99}, is perfectly known \cite{John04,ACPRA2009}.}.
Further investigations suggested to add external magnetic fields to open band gaps in other geometric structures \cite{YUPRA2011}.

In cold atom realizations of $3\mathrm{D}$ optical lattices, the atomic Mott state extends over $10-20$ lattice sites, so a natural question regards the features of the band gap in this finite size system. A further question concerns the effects of an imperfect finite portion of a lattice containing site defects, i.e.\ a fraction of vacancies resulting in a not complete filling of the lattice. The experimental interest of these issues is related to the fact that both the finite size and vacancy effects, separately, could in principle drastically affect the presence and the experimental visibility of the band gap due to the appearance of states in the gap region. The main  questions we address in this paper are: What does happen to the band gap for systems of realistic sizes and of different shapes? What is the fraction of vacancies which still permit to have a reasonable band gap visibility? What is the value of the penetration depth of an electromagnetic wave in the atomic diamond lattice for finite and infinite systems, i.e.\ how is it affected by finite size effects?  Even if we discuss in detail the case of a diamond lattice, we will present a general formulation and will discuss main features which will remain valid for other lattice geometries.   

The paper is organized as follows. In section \ref{sec:model} we illustrate the model we use, and the resulting main equations we solve. In section \ref{sec:num} we present and discuss a numerical study on the density of states and on the penetration length in a finite size system, possibly in presence of imperfections due to vacant sites in the lattice. In section \ref{sec:the} we provide an analytical analysis to support and illustrate some of the main features of the numerical findings. We conclude in section \ref{sec:conclusion}.

\section{The model \label{sec:model}}
We consider a system made by a collection of $N$ identical atoms having fixed positions and an optical dipolar transition between a $J_g=0$ electronic ground state and a $J_e=1$ electronic excited state \cite{scalar}. Such a transition is available in appropriate atomic species,  such as
strontium where it was already used to study coherent propagation of light in an atomic ensemble \cite{Wilk}.
In our model, the atomic dipoles are coupled by the electromagnetic field they radiate, and  in the regime of low atomic excitations, the resulting eigenmodes of the mean atomic dipoles are given by the solutions of the eigenvalue problem \cite{Morice95,Coevorden96}
\begin{multline} \label{eq:dipoli}
\left(\hbar\omega_{0}-i\frac{\hbar\Gamma}{2}\right)d_{i,\alpha}+\sum_{\substack{j=1 \\ j\neq i}}^{N}\sum_{\beta=x,y,z}g_{\alpha\beta}(\rr_i-\rr_j)d_{j,\beta}=\\
\hbar(\omega-i\gamma) d_{i,\alpha}.
\end{multline}
Here $d_{i,\alpha}$ is the component along the direction $\alpha=x,y,z$ of the mean dipole carried by the atom $i$, $\omega-i\gamma$ is the mode eigenfrequency (it is complex in general with $\gamma>0$, and may be measured as suggested in \cite{ACPRL2009}), $\omega_0$ and $\Gamma$ are the single atom resonance frequency and spontaneous emission rate.
The tensor $g_{\alpha\beta}(\rr)$ gives the $\alpha$ component of the electric field at the position $\rr$ radiated by a dipole oscillating along the direction $\beta$ at the origin of coordinates, $E_{\alpha}(\rr)=-g_{\alpha \beta}(\rr)d_{\beta}/\mathrm{d}^2$,  $\mathrm{d}$ being the atomic dipole moment such that $\Gamma=\mathrm{d}^2\omega_0^3/(3\pi\varepsilon_0\hbar c^3)$. Here we consider the case where $\omega-i\gamma$ is very close to $\omega_0$, so that $g_{\alpha \beta}$ can be evaluated for a dipole oscillating at the resonance frequency; introducing the vacuum wavenumber 
\be
\label{eq:defk0}
k_0=\frac{\omega_0}{c}
\ee
we take 
\begin{multline}\label{eq:g}
g_{\alpha\beta}(\rr)  = -\frac{3\hbar\Gamma}{4k_0^3}[(k_0^2\delta_{\alpha\beta}+\partial_{r_\alpha}\partial_{r_\beta})
\frac{e^{ik_0r}}{r}+4\pi\delta_{\alpha\beta}\delta(\rr)] \\  \underset{r>0}{=}\frac{3}{4}\hbar\Gamma\;\frac{e^{ik_0r}}{k_0r}
\left[\left(-1-\frac{i}{k_0r}+\frac{1}{(k_0r)^2}\right)\delta_{\alpha\beta} \right.\\
+\left.\left(1+\frac{3i}{k_0r}-\frac{3}{(k_0r)^2}\right)\frac{r_{\alpha}r_{\beta}}{r^2}\right].
\end{multline}
Our first expression in (\ref{eq:g}) for $g_{\alpha\beta}(\rr)$ differs by a scalar $\delta(\rr)$ contribution from
the usual expression for the electric field radiated by a dipole, see Eqs.~(4.20,9.18) of \cite{Jackson}; this ensures compatibility
with our previous works and it is of course irrelevant here since atoms are never at the same position
\footnote{Our convention amounts to omitting the $\delta(\rr)$ term in Eq.~(3) of \cite{Morice95}.}. The first expression is particularly useful to directly extract its Fourier transform, needed in the Bloch-description of infinite systems (see section \ref{sec:the}), while the second one (which differs from the first one by another scalar $\delta(\rr)$ contribution) has the well know dipole-dipole interaction form, and will be used in numerical calculations on finite-size systems in section \ref{sec:num}.

Equation  (\ref{eq:dipoli}) allows one to determine the density of states of the system. In case an infinite number of atoms are periodically arranged at the nodes of a diamond lattice, it has been shown that the system may exhibit
 an omnidirectional photonic band gap \cite{ACPRA2009}. Here, by numerical solution of (\ref{eq:dipoli})
 we investigate the fate of such a gap, in situations close to realistic experimental ones, where the number of atoms is finite and/or there are vacancies in the lattice. A further interesting quantity related to the occurrence of a gap is the so-called ``penetration depth" $\xi$: an incident electromagnetic wave at a frequency in the band gap cannot penetrate the medium, and its amplitude will decay exponentially over a characteristic distance $\xi$.  In order to calculate such a length we consider a point-like dipolar source immersed in the atomic medium, and we extract $\xi$ from the total field and the induced dipole spatial profiles: we fix at the position $\rr_s$ a forced dipole $d_\alpha^{\mathrm s}=\check{d}_\alpha^{\mathrm s}\,e^{-i\omega_s t}$, the atomic dipoles at the positions $\rr_i$ will reach a steady state $d_{i,\alpha}=\check{d}_{i,\alpha}\,e^{-i\omega_s t}$ given by the linear system

\begin{multline}\label{eq:steady}
-\left[\hbar(\omega_s-\omega_{0})+i\frac{\hbar\Gamma}{2}\right]\check{d}_{i,\alpha}+\sum_{\substack{j=1 \\ j\neq i}}^{N}\sum_{\beta=x,y,z}g_{\alpha\beta}(\rr_i-\rr_j)\check{d}_{j,\beta}=\\
-\sum_{\beta=x,y,z}g_{\alpha\beta}(\rr_i-\rr_s)\check{d}^{s}_{\beta}.
\end{multline}

\section{Numerical results for a finite size system \label{sec:num}}
In this section we study a system of $N$ atoms at the nodes of a diamond lattice. We recall that the diamond lattice is formed
by the superposition of two copies of the same Bravais lattice: the fcc lattice
of lattice constant $a$, generated by the three basis vectors
\be
\label{eq:ei}
\mathbf{e}_1=(0,a/2,a/2), \mathbf{e}_2=(a/2,0,a/2), \mathbf{e}_3=(a/2,a/2,0),
\ee
 and a second fcc lattice obtained by translating the first lattice by the vector $(a/4,a/4,a/4)$.
 The corresponding basis of the reciprocal lattice is 
\begin{multline}
\label{eq:tei}
\tilde{\eee}_1=(-2\pi/a,2\pi/a,2\pi/a),
\tilde{\eee}_2=(2\pi/a,-2\pi/a,2\pi/a),  \\
\tilde{\eee}_3=(2\pi/a,2\pi/a,-2\pi/a).
\end{multline}
In our simulations, the atoms occupy a finite region in space, which can be a ball or a cube centered at the origin of the coordinates. From the numerical solution of (\ref{eq:dipoli}) we extract the density of states for the case of a unit filling factor (section \ref{ss:fse}) and for the case with a low concentration of vacancies (section \ref{ss:vac}).  Finally, we analyze the penetration depth in section \ref{ss:pd} solving (\ref{eq:steady}). 

\subsection{Finite size effects on the density of states \label{ss:fse}}
\begin{figure}[tb]
\includegraphics[width=\columnwidth,clip=]{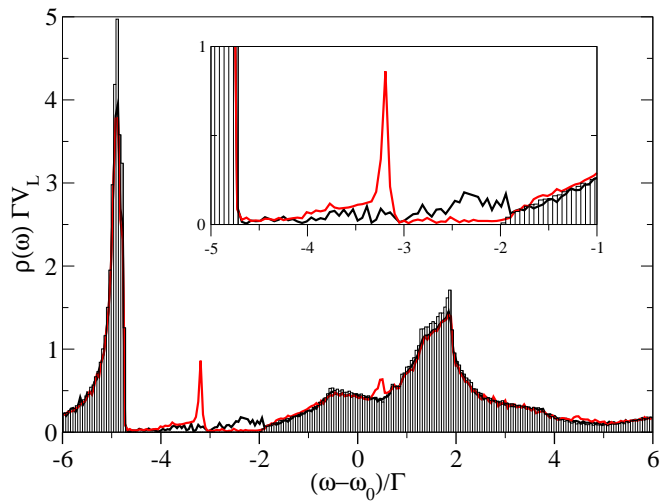}
\caption{(Color online) Density of the real part of the eigenfrequencies $\rho(\omega)$ obtained from Eq.(\ref{eq:dipoli}), in the absence of vacancies and for $k_0a=2$ where $a$ is the fcc lattice constant. Red solid lines: finite system with a cubic shape of side of length $14a$, and $N=2.7\times10^4$. Black solid lines: finite system with a spherical shape of diameter $18a$, and $N=2.4\times10^4$. The histogram provides the same quantity for an infinite system \cite{ACPRA2009}. Each of the three curves is composed of 250 bins. $\mathcal{V}_{\mathrm{L}}$ is the volume of the  direct lattice primitive cell. The inset is a magnification. \label{fig:dos}}
\end{figure}

In this section we discuss the density of states obtained by solving equation (\ref{eq:dipoli}) for a finite size diamond lattice, in the absence of vacancies.  
In particular, in Fig.~\ref{fig:dos} we show the density of states $\rho(\omega)$ for a number of atoms corresponding to typical experimental realizations $N\approx2.5\times10^4$. Here $\rho(\omega)$ is defined as the distribution of the real part $\omega$ of the complex spectrum of equation (\ref{eq:dipoli}), normalized as $\int \rho(\omega) d\omega=6/\mathcal{V}_{\mathrm{L}}$, where $\mathcal{V}_{\mathrm{L}}=a^3/4$ is the volume of the  direct lattice primitive cell, in order to facilitate the comparison with the infinite system results of \cite{ACPRA2009}, plotted as a bar histogram in the figure. If the atoms occupy a ball (see the black solid line) we observe partial filling of the spectral gap, most pronounced in the upper region. On the contrary the region close to the lower border of the gap remains relatively weakly affected by the finite size of the system, considering the sharp rise of $\rho(\omega)$ to the left of this border. The remaining part of the density of states remains very close to the one of the infinite system. 
If the atoms occupy a cube (see the red solid line) the finite size effects are quite different. Two peaks appear, a very pronounced one in the middle of the gap (at $(\omega-\omega_0)/\Gamma\approx-3.2$), and a second one (at $(\omega-\omega_0)/\Gamma\approx0.5$). We investigated the nature of the states belonging to the peak in the gap, by looking at $10$ successive eigenstates of (\ref{eq:dipoli}), finding that they are ``border states'': they reach their maxima in a spherical shell of radius $\approx5a$, and decay exponentially towards the center of the cube with a law 
\be
|d_i|^2\equiv \sum_{\alpha=x,y,z} |d_{i,\alpha}|^2 \lesssim  e^{-22+4.6 r_i/a} 
\ee
 where the dipole eigenvectors are normalized to the maximum value of their modulus equal to unity. This suggests a value of the penetration depth of  the order of $0.5a$, in agreement with the calculation done in section \ref{ss:pd}. 

\begin{figure}[tb]
\includegraphics[width=\columnwidth,clip=]{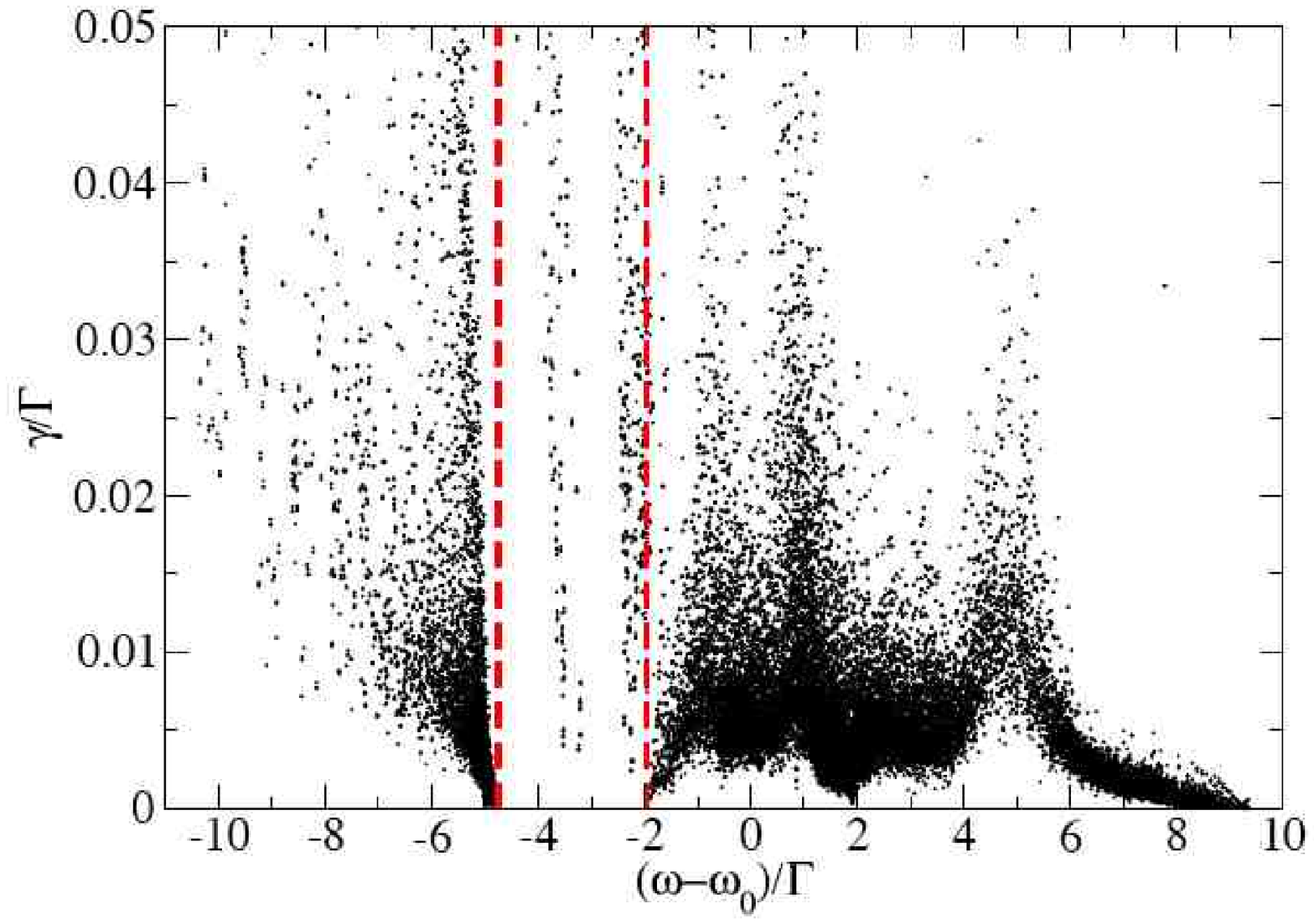}
\caption{(Color online) Complex eigenvalues $\omega-i\gamma$ obtained from Eq.(\ref{eq:dipoli}). The system is of finite size,  in the absence of vacancies, for $k_0a=2$, with a spherical shape of diameter $18a$, and $N=2.4\times10^4$ atoms. The two vertical red dashed lines give the borders of the band gap of the infinite periodic system. \label{fig:spectrum01}}
\end{figure}

In Fig.~\ref{fig:spectrum01}  we show the distribution of the eigenvalues of Eq.(\ref{eq:dipoli}) in the complex plane, restricted to small values of $\gamma/\Gamma$. In this region, the figure shows that the band gap is not filled, apart from two narrow intervals of values of $\omega$, in the center and close to the upper border of the gap.  Then, in the finite size system, the partial  filling of the gap is mostly due to eigenvalues with larger values of $\gamma/\Gamma$. The smallest values 
of $\gamma/\Gamma$ we obtained are $\approx 2\times 10^{-5}$. The real part of the corresponding eigenvalues are located on the borders of the band gap for the infinite system, marked in the figure by vertical dashed lines, and on the upper bound of the values of $\omega$ represented in Fig.~\ref{fig:spectrum01} for the finite system, i.e.\ around $(\omega-\omega_0)/\Gamma=9.5$.

\subsection{Effects of vacancies on the density of states \label{ss:vac}}
\begin{figure}[tb]
\includegraphics[width=\columnwidth,clip=]{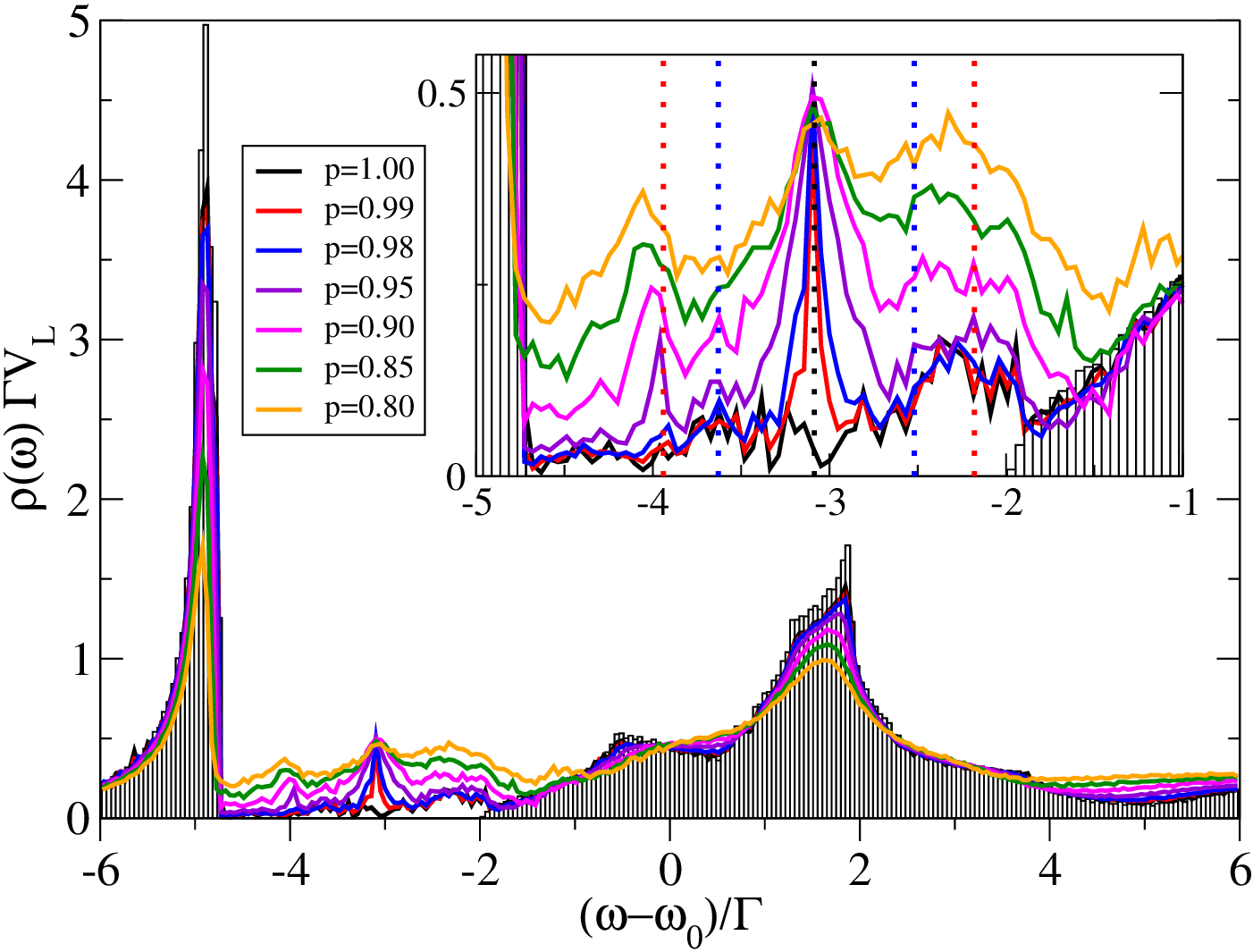}
\caption{(Color online) Density of the real part of the eigenfrequencies $\rho(\omega)$ obtained from Eq.(\ref{eq:dipoli}), for different concentrations of vacancies, that is for various filling factors $p$, and for $k_0a=2$. The finite system has a spherical shape of diameter $18a$, and $N=2.4\times10^4$ for $p=1$. The histogram provides the same quantity for an infinite system with no vacancies \cite{ACPRA2009}.  $\mathcal{V}_{\mathrm{L}}$ is the volume of the  direct lattice primitive cell. The
inset is a magnification, where the vertical dotted 
lines correspond to frequencies of the single vacancy in-gap state (black, central) and to two-vacancy in-gap states ($\breve{R}_2-\breve{R}_1=\eee_1$,
$\breve{\mu}_1=\breve{\mu}_2=1$ in red, outer; $\breve{R}_2-\breve{R}_1=a\eee_x$, $\breve{\mu}_1=2, \breve{\mu}_2=1$ in blue, inner; 
these quantities are defined in appendix~\ref{app:sic}) theoretically predicted in section \ref{subsec:vac}. Decreasing values of $p$ correspond to increasing values of $\rho(\omega)$ in the band gap. \label{fig:dosvac}}
\end{figure}

In this section we address the case where the finite size diamond lattice in not perfectly filled, presenting a concentration $1-p$ of defects made by the presence of a random uniform distribution of not-occupied lattice sites. In Fig.~\ref{fig:dosvac} we show the density of states $\rho(\omega)$ obtained solving Eq.~(\ref{eq:dipoli})  for atoms occupying a ball, as a function of the lattice filling factor $p$. The figure, and its inset, show that already a small concentration of vacancies equal to $1-p=0.99$  (red solid line) produces a remarkable signature in the density of states manifested by the appearance of a pronounced peak in the middle of the band gap, at $(\omega-\omega_0)/\Gamma\approx-3.08$. We explain the nature of the peak with the presence of single-vacancy states localized at the vacancy position. Since the vacancy concentration is small, most frequent vacancy states have a single-site nature. In section  \ref{subsec:vac} we theoretically calculate the value of the single-vacancy state frequency, signaled in the inset by  a black vertical dotted line, which seems to coincide quite satisfactorily with that of the numerically observed peak. 
By increasing the vacancy concentration, Fig.~\ref{fig:dosvac} shows for $1-p=0.05$ the occurrence of a clear second peak in the gap, 
which seems to match quite well the frequency of a two-vacancy in-gap state calculated in section  \ref{subsec:vac}, see the red vertical dotted line at
$(\omega-\omega_0)/\Gamma \simeq -4$. Peaks corresponding to other two-vacancy states predicted in section  \ref{subsec:vac} are less visible
(see the other vertical dotted lines in the inset of Fig.~\ref{fig:dosvac}).
Further increase of the concentration of vacancies produces a gradual filling of the band gap, whose visibility completely deteriorates for a vacancy concentration around $1-p=0.2$. 

\begin{figure}[tb]
\includegraphics[width=\columnwidth,clip=]{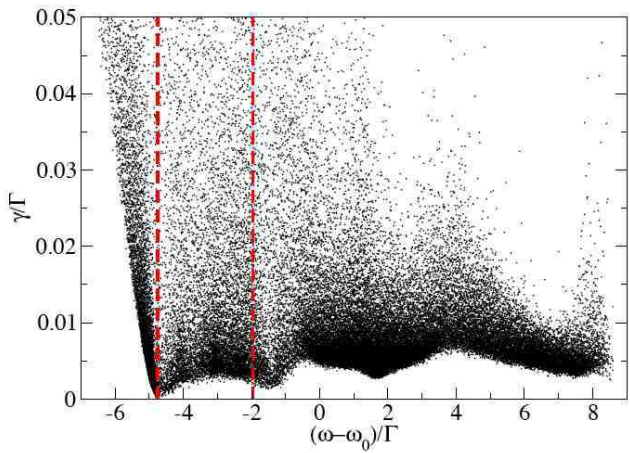}
\caption{(Color online) Complex eigenvalues $\omega-i\gamma$ obtained from Eq.(\ref{eq:dipoli}). The system is of finite size, in presence of vacancies, that is with a filling factor $p=0.8$, for $k_0a=2$, with a spherical shape of diameter $18a$, and $N=1.9\times10^4$ atoms. The two vertical red dashed lines give the borders of the band gap of the infinite periodic system. \label{fig:spectrum08}}
\end{figure}

In Fig.~\ref{fig:spectrum08}, for exactly the same spherical system with a vacancy concentration of $1-p=0.2$, we show  the distribution of the eigenvalues of Eq.(\ref{eq:dipoli}) in the complex plane, restricted to small values of $\gamma/\Gamma$. The figure shows that the band gap is completely filled. The states filling the gap, for such a large vacancy concentration, are completely delocalized over the entire system size, and have a spectral imaginary part mostly concentrated around $\gamma/\Gamma\approx10^{-2}$, with $\gamma/\Gamma\ge10^{-3}$.

\begin{figure}[tb]
\includegraphics[width=\columnwidth,clip=]{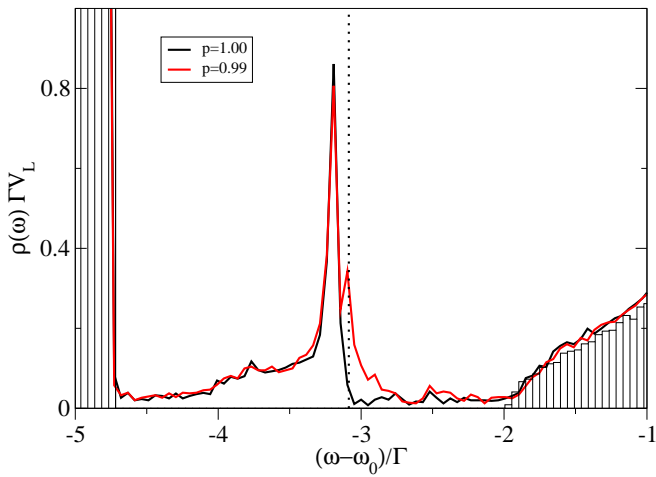}
\caption{(Color online) Density of the real part of the eigenfrequencies $\rho(\omega)$ obtained from Eq.(\ref{eq:dipoli}), for two concentrations of vacancies, that is for the filling factors $p=1$ and $p=0.99$, and for $k_0a=2$. The finite system has a cubic shape of side $14a$, and $N=2.7\times10^4$ for $p=1$. The histogram provides the same quantity for an infinite system with no vacancies \cite{ACPRA2009}.  $\mathcal{V}_{\mathrm{L}}$ is the volume of the  direct lattice primitive cell. We note the double peak structure for $p=0.99$ (see text). The vertical dotted line corresponds to the frequency of the single vacancy in-gap state theoretically predicted in section  \ref{subsec:vac}. \label{fig:dosvaccube}}
\end{figure}

In Fig.~\ref{fig:dosvaccube} we study the effect of vacancies on a system of cubic shape. The figure shows that for a concentration of vacancies $1-p=0.01$ (red solid line) two peaks are present in the band gap.
They have a different origin: the first one, that at smallest energy, in nothing but the peak related to shape-induced states, already present in the absence of vacancies (see black solid line, and the discussion in section \ref{ss:fse}). The second peak is instead the signature of single-vacancy localized states, and its position is the same of that shown in Fig. \ref{fig:dosvac} for spherical shape at the same vacancy concentration.  

\subsection{Penetration depth \label{ss:pd}}

To numerically calculate the penetration depth $\xi$ for a diamond finite-size atomic lattice we numerically solve the forced dipole equation (\ref{eq:steady}) for a point-like dipolar oscillating source $d_\alpha^{\mathrm s}=\check{d}_\alpha^{\mathrm s}\,e^{-i\omega_s t}$ at the position $\rr_s$ (approximately at the center of the system), and with $\omega_s$ in the band gap. Solutions of Eq.(\ref{eq:steady}) provide the  induced atomic dipoles amplitudes $\check{d}_{i,\alpha}$ at the lattice positions $\rr_i$. 

\begin{figure}[tb]
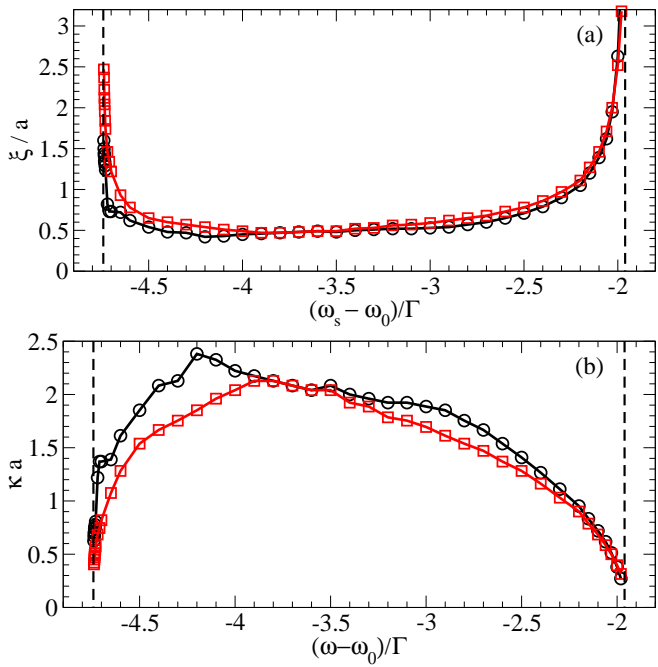

\includegraphics[width=\columnwidth,clip=]{fig6a.eps}
\includegraphics[width=\columnwidth,clip=]{fig6b.eps}
\caption{(Color online) Penetration depth $\xi$ in (a) and its inverse $\kappa$ in (b), as functions of the dipole source
frequency $\omega_s$. Symbols (the lines are a guide to the eye) correspond to the numerical solution of Eq.(\ref{eq:steady}) for a finite system of spherical shape, diameter $18a$, filling factor $p=1$, $k_0a=2$, and containing $N=2.4\times10^4$ atoms. Red squares and black circles correspond to values obtained using the methods of Eq.(\ref{eq:pendepd}) and of Eq.(\ref{eq:pendepf}), respectively.  The vertical dashed lines corresponds to the borders (\ref{eq:borper})  of the band gap for the infinite periodic system. 
\label{fig:pendep}}
\end{figure}

\begin{figure}[tb]
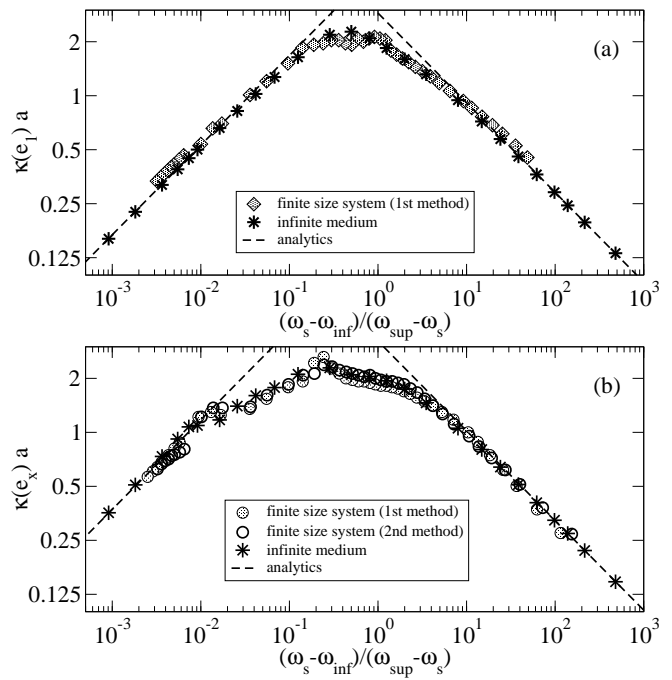

\includegraphics[width=\columnwidth,clip=]{fig7a.eps}
\includegraphics[width=\columnwidth,clip=]{fig7b.eps}
\caption{Inverse penetration depth $\kappa=1/\xi$ along direction $\eee_1$ in (a) and direction $\eee_x$ in (b),
as a function of the source frequency $\omega_s$ 
expressed through a change of variable mapping the band gap $[\omega_{\rm inf},\omega_{\rm sup}]$ onto $\mathbb{R}^+$. Same physical 
parameters as in Fig.~\ref{fig:pendep}. Filled diamonds in (a) and filled circles in (b): finite size system with first extraction method; 
empty circles in (b): finite size system with second extraction method;
for those data, $\omega_{\rm inf}$ and $\omega_{\rm sup}$ were obtained by linear extrapolation of $\kappa^2$ to zero.
Stars: for the infinite medium from a numerical evaluation of Eq.~(\ref{eq:dip_force}).
Dashed lines: analytical predictions, close to the band borders, deduced from Eq.~(\ref{eq:kappa}) (see text).
Note that the $x$ and $y$ axes are in $\log_{10}$ and $\log_{2}$ scale.
\label{fig:pendepzoom}}
\end{figure}

We extract $\xi$ according to different methods. 
The first method is based on the direct analysis of the induced dipoles, and consist in averaging the norm $\sqrt{\sum_{\alpha}|\check{d}_{i,\alpha}|^2}$ on spherical shells of radius $\approx u=||\rr-\rr_s||$ centered at the source position. We then obtain an average real dipole function $\mathcal{D}(u)$ that we fit  in a certain range of $u$ (where the behavior of $d(u)$ is clearly exponential over several decades) as 
\be\label{eq:pendepd}
\mathcal{D}(u)=C\;\frac{e^{-u/\xi}}{u}
\ee
where $\xi$ and $C$ are the two fitting parameters. The factor $1/u$ in (\ref{eq:pendepd}) is introduced to take into account the direct effect of the source which is dominant at small distances, allowing to fit the function on a larger range. This method provides the results presented by red squares in Fig.~\ref{fig:pendep}.
Its specialisation to the analysis of the penetration depth along some given direction (without averaging over
spherical shells) is straightforward, and leads to the filled diamonds and circles in Fig.~\ref{fig:pendepzoom}a and b, respectively.

The second method is based on the calculation of the total electric field amplitude generated by the source and induced dipoles obtained by (\ref{eq:steady}) :  
\be\label{eq:ef}
\check{E_{\alpha}}(\rr)=-\sum_{\beta}g_{\alpha\beta}(\rr-\rr_s)\frac{\check{d}_{\beta}^{\mathrm s}}{\mathrm{d}^2}-\sum_{i=1}^N\sum_{\beta}g_{\alpha\beta}(\rr-\rr_i)\frac{\check{d}_{i,\beta}}{\mathrm{d}^2}, 
\ee
We evaluate $\check{E_{\alpha}}(\rr)$ on three lines, parallel to the Cartesian axes and passing trough the source position $\rr_s$. We first average the norm $\sqrt{\sum_{\alpha}|\check{E}_{\alpha}(\rr)|^2}$ on the two directions $(\pm)$ of the three axes $\alpha$, then we obtain and fit the six corresponding average real electric functions $\mathcal{E}_{\alpha}^{(\pm)}(u)$ as
\be\label{eq:pendepf}
\mathcal{E}_{\alpha}^{(\pm)}(u)=K_{\alpha}^{(\pm)}\;\frac{e^{-u/\xi_{\alpha}^{(\pm)}}}{u},
\ee
obtaining six values of $\xi_{\alpha}^{(\pm)}$, whose average is presented by empty black circles in Figs.~\ref{fig:pendep}
and \ref{fig:pendepzoom}b.

In Fig.~\ref{fig:pendep}, it is apparent that the extractions of the penetration depth from Eq.~(\ref{eq:pendepd})
and from Eq.~(\ref{eq:pendepf}) give different values. 
This shows that $\xi$ is not isotropic, it depends on the considered direction of space, a property
that will be recovered analytically in section \ref{ss:pen}.
Whereas use of Eq.~(\ref{eq:pendepf}) is expected to give the penetration depth along $x$ axis, the first method, when it involves 
a directional average as in Eq.~(\ref{eq:pendepd}), is expected to pull out the maximal penetration depth (maximized over the directions of space).
A second property, apparent in Fig.~\ref{fig:pendep}a, is the divergence of $\xi$ at the borders of the infinite-medium forbidden gap
(represented by vertical dashed lines at frequencies $\omega_{\rm inf}$, $\omega_{\rm sup}$). 
Fig.~\ref{fig:pendep}b even suggests that $\kappa$ vanishes there with a vertical slope. We indeed
find that $\kappa^2$ vanishes linearly with $\omega_s$ (not shown), as also predicted
analytically in section \ref{ss:pen}. By a linear extrapolation of $\kappa^2$ as a function of $\omega_s$, we get
for the borders of the forbidden bands:
\bea
\label{eq:bords_noir}
(\frac{\omega_{\rm inf}-\omega_0}{\Gamma},\frac{\omega_{\rm sup}-\omega_0}{\Gamma})\!\! &\stackrel{\mathrm{Eq.~(\ref{eq:pendepf})}}{\simeq}&
\!\!  (-4.748,-1.962) \\
\label{eq:bords_rouge}
\!\!&\stackrel{\mathrm{Eq.~(\ref{eq:pendepd})}}{\simeq}& \!\! (-4.747,-1.948)
\eea
which are indeed quite close to the infinite medium results \cite{ACPRA2009}:
\be
\label{eq:borper}
(\frac{\omega_{\rm inf}-\omega_0}{\Gamma},\frac{\omega_{\rm sup}-\omega_0}{\Gamma}) \simeq (-4.743, -1.962).
\ee

To better put in evidence the vanishing of $\kappa$ at the band edges, and to more easily compare
the various methods, we show $\kappa$ as a function
of $(\omega_s-\omega_{\rm inf})/(\omega_{\rm sup}-\omega_s)$ in Fig.~\ref{fig:pendepzoom}, with the band edges $\omega_{\rm inf}$ and
$\omega_{\rm sup}$ deduced for the finite-size simulations by linear extrapolation of $\kappa^2$
\footnote{Note that, according to Eqs.~(\ref{eq:relgeni},\ref{eq:relsour}) to come, this rational fraction of the source frequency
is the same for the original model and the Gaussian spatially smoothed model, 
$(\bar{\omega}_s-\bar{\omega}_{\rm inf})/(\bar{\omega}_{\rm sup}-\bar{\omega}_s)
\simeq (\omega_s-\omega_{\rm inf})/(\omega_{\rm sup}-\omega_s)$, within an exponentially small error in $1/b^2$.}.
This change of variable on $\omega_s$ has the advantage of mapping the band edges to $0$ and $+\infty$,
respectively, which is then combined with a log-scale representation on both figure axes.
This figure was produced for two particular directions of penetration, along the direct lattice basis vector
$\eee_1$ in Fig.~\ref{fig:pendepzoom}a, and along the Cartesian axis direction $\eee_x$ in Fig.~\ref{fig:pendepzoom}b.
First, in Fig.~\ref{fig:pendepzoom}b, it appears that the two extraction methods for the penetration depth in the finite-size
system (the first method from the dipoles, see the filled circles; the second method from the electric field, see the empty
circles) give compatible results if they are applied
along the {\sl same} direction (here $\eee_x$, which is equivalent to $\eee_y$ or $\eee_z$ due to symmetry
of the diamond lattice). Second, in Fig.~\ref{fig:pendepzoom}a and b, the results of the finite-size systems are compatible with the ones (stars) for the infinite
system in section \ref{ss:pen}, and even if they do not cover a as large range for $\kappa$, they do nicely follow
the analytical prediction (dashed lines) for the vanishing of $\kappa$ close to the band edges.

\section{Theory for the infinite system \label{sec:the}}

We show in this section that several features of the numerical simulations, such as the sharp rise of $\xi$ close to the band gap
borders and some peaks induced by vacancies in $\rho(\omega)$, can be interpreted analytically for an infinite system.
In this case, a reformulation of (\ref{eq:dipoli},\ref{eq:steady}) in Fourier space is more appropriate. It is known however that
the resulting series over the reciprocal lattice present subtle convergence issues \cite{Knoester06} that were
overlooked in \cite{Coevorden96,LagendijkRMP}. 
These issues were solved in \cite{ACPRA2009} by coupling each atomic dipole to a spatially smoothed
version $\bar{\mathbf{E}}_\perp(\rr)= \int d^3u \mathbf{E}_\perp(\rr-\uu) \chi(\uu)$  of the transverse electromagnetic field operator $\mathbf{E}_\perp(\rr)$,
where the smoothing function $\chi(\uu)$ may be taken as a positive rotationally invariant function of unit integral and of small width $b$.
This cuts off the dipolar coupling at high wavenumber field modes and regularizes the theory for the infinite system. 

One then finds that two changes have to be applied to Eqs.~(\ref{eq:dipoli},\ref{eq:steady}).
First, the function $g_{\alpha\beta}(\rr)$ has to be replaced by the smoothed function $\bar{g}_{\alpha\beta}(\rr)$ such that
\be
\label{eq:gbar}
\bar{g}_{\alpha\beta}(\rr_i-\rr_j) =\!\!\int\!\! d^3 u_i\! \int\!\! d^3 u_j g_{\alpha\beta}(\rr_i+\uu_i-\rr_j-\uu_j) \chi(\uu_i) \chi(\uu_j).
\ee
In Fourier space, the convolution products take a simple form so that
\be
\label{eq:gbarf}
\tilde{\bar{g}}_{\alpha\beta}(\kk)=\frac{3\pi\hbar\Gamma}{k_0^3}\,\frac{k^2\delta_{\alpha\beta}-k_{\alpha}k_{\beta}}{k_0^2-k^2+i0^+} \tilde{\chi}^2(\kk)
\ee
where $\tilde{\bar{g}}_{\alpha\beta}(\kk)=\int d^3r\, e^{-i\kk\cdot\rr} \bar{g}_{\alpha\beta}(\rr)$ is the Fourier transform of $\bar{g}_{\alpha\beta}$
and $\tilde{\chi}(\kk)$ is the one of  $\chi(\rr)$.
Second, the spontaneous emission rate $\Gamma$ in Eqs.~(\ref{eq:dipoli},\ref{eq:steady}) has to be replaced by
\be \label{eq:Gamma_bar}
\bar{\Gamma} = \Gamma \tilde{\chi}^2(\kk_0)
\ee
where $\kk_0$ is a vector of modulus equal to $k_0$ and of arbitrary direction.
If one would treat the atomic motion quantum mechanically, as in \cite{ACPRL2009}, for atoms trapped at the nodes of an optical lattice,
$\chi(\uu)=\phi^2(\uu)$ would be the probability distribution of the fluctuations $\uu$ of the atomic position around a node $\rr_i$, where $\phi$ is the underlying
atomic center-of-mass wavefunction. Then Eq.~(\ref{eq:gbar}) would have a straightforward physical interpretation. 
Also $\bar{\Gamma}$ would simply be the elastic spontaneous emission
rate, where the atomic center-of-mass after decay to the electronic ground state remained in the wavefunction $\phi$. 
In practice, a Gaussian choice for $\chi$  is convenient, which corresponds to 
\be 
\label{eq:cogauss}
\tilde{\chi}(\kk)=e^{-k^2b^2/2}.
\ee

It is useful to know to which extent the results from the spatially smoothed model
differ from the original model. For the Gaussian smoothing function, one then has the remarkable result that, when 
the width $b$ is much smaller than all interatomic distances $|\rr_i-\rr_j|$, one has the approximate
relation
\be
\bar{g}_{\alpha\beta}(\rr_i-\rr_j) \simeq e^{-k_0^2 b^2} g_{\alpha\beta}(\rr_i-\rr_j)
\ee
with an exponentially small error in $1/b^2$ \cite{ACPRL2009}, that is one has the same Gaussian
factor as for $\bar{\Gamma}$.
For the eigenvalue problem (\ref{eq:dipoli}), this shows that the eigenvalues $\bar{\omega}-i\bar{\gamma}$
of the spatially smoothed model may be related to the ones $\omega-i\gamma$ of the original model by
\be
\label{eq:relgeni}
\bar{\omega}-\omega_0-i\bar{\gamma} \simeq e^{-k_0^2 b^2} (\omega-\omega_0-i\gamma)
\ee
within an exponentially small error in $1/b^2$.  For the steady state problem (\ref{eq:steady}), it is found
that the forced dipoles of the spatially smoothed model will (within an exponentially small error) coincide
with the ones of the original model if one takes in the smoothed model the modified source frequency such that
\be
\label{eq:relsour}
\bar{\omega}_s-\omega_0 = e^{-k_0^2 b^2} (\omega_s-\omega_0).
\ee

\subsection{Density of states for the infinite periodic system \label{ss:dosps}}
In this subsection, we show how to recover Fourier space results  of \cite{ACPRA2009} for the density of states in the infinite periodic system, starting from
the smoothed version of the real space Eq.~(\ref{eq:dipoli}). 

According to Bloch theorem, solutions of (\ref{eq:dipoli}) can be taken of
the form $d_{i,\alpha}=d_\alpha^{(\mu)}\;e^{i\qq\cdot\RR}$, where $\qq$ is the Bloch vector, $\RR$ is a vector of the Bravais lattice, the index $\mu$ labels  primitive cells (for the diamond lattice, given by the combination of two shifted fcc Bravais lattices, $\mu$ assumes two values), so that all atomic positions can be written as $\rr_i=\RR+\rr^{(\mu)}$, where $\rr^{(\mu)}$ is the position with respect to the  Bravais lattice vector $\RR$. Injecting this ansatz in Eq.~(\ref{eq:dipoli}) modified according
to Eqs.~(\ref{eq:gbar},\ref{eq:Gamma_bar}), gives the eigenvalue problem
\be\label{eq:eigb}
\sum_{\beta,\nu} \bar{\mathbb{P}}_{\alpha\mu,\beta\nu}(\qq) \bar{d}_\beta^{(\nu)} = \hbar(\bar{\omega}-\omega_0-i\bar{\gamma}) 
\bar{d}_\alpha^{(\mu)}
\ee
with
\begin{multline}
\label{eq:Pini}
 \bar{\mathbb{P}}_{\alpha\mu,\beta\nu}(\qq) = -\left[\bar{g}_{\alpha\beta}(\mathbf{0})+i \frac{\hbar\bar{\Gamma}}{2} \delta_{\alpha\beta}\right]\delta_{\mu\nu}\\
 +\sum_{\RR\in \mathrm{L}} \bar{g}_{\alpha\beta}(\RR+\rr^{(\mu)}-\rr^{(\nu)})e^{-i\qq\cdot\RR}.
\end{multline}
Here indices $\alpha,\beta$ and $\mu,\nu$ label the direction and primitive cell, respectively, and eigenvalues $\bar{\omega}-i\bar{\gamma}$ and eigenvectors $\bar{d}_\beta^{(\nu)}$ depend on the choice of the cut-off smooth function $\chi(\uu)$, hence for the Gaussian
choice as in Eq.(\ref{eq:cogauss}), they depend on the value of $b$.
By considering the first contribution of Eq.(\ref{eq:Pini}), inside the square brackets,  it is found from the inverse Fourier transform of (\ref{eq:gbarf}) 
that the tensor $\bar{g}_{\alpha\beta}(\mathbf{0})$ is scalar (it is proportional
to $\delta_{\alpha\beta}$); further, using $1/(X+i0^+)=\mathcal{P}\frac{1}{X}-i\pi\delta(X)$ and (\ref{eq:Gamma_bar}), one finds that the imaginary part of 
$\bar{g}_{\alpha\beta}(\mathbf{0})$ exactly cancels with the $\bar{\Gamma}$ term. The second contribution, that is the sum over the Bravais lattice
in (\ref{eq:Pini}), can be transformed with the Poisson summation formula.  For the Gaussian smoothing function (\ref{eq:cogauss}), the real part of
$\bar{g}_{\alpha\beta}(\mathbf{0})$ can be calculated explicitly; one obtains as in \cite{ACPRA2009}:
\begin{multline}
\label{eq:cas_pertu}
\bar{\mathbb{P}}_{\alpha\mu,\beta\nu}(\qq) =  \frac{\hbar\Gamma}{2} \delta_{\alpha\beta} \delta_{\mu\nu}
\left[\frac{1+2(k_0b)^2}{2\pi^{1/2} (k_0b)^3}-\textrm{Erfi}\,(k_0b) e^{-k_0^2b^2}\right]
\\
+ \frac{1}{\mathcal{V}_\textrm{L}}
\sum_{\KK\in \textrm{RL}} e^{i(\KK+\qq)\cdot(\rr^{(\mu)}-\rr^{(\nu)})}
\tilde{\bar{g}}_{\alpha\beta}(\KK+\qq)
\end{multline}
where the wavevectors $\KK$ run over the reciprocal lattice of the Bravais lattice, and $\mathrm{Erfi}$ is the imaginary
error function. As expected for an infinite system, the matrix $\bar{\mathbb{P}}$ is hermitian,
so that $\bar{\gamma}=0$. 

Turning back to the original problem (\ref{eq:dipoli}), that is in the absence of any smoothing function,
we conclude for the infinite periodic system that the spectrum is real ($\gamma=0$) and that $\hbar\omega-\hbar\omega_0$ is any of the eigenvalues
of the matrix 
\be
\mathbb{P}(\qq)=\lim_{b\to 0} \bar{\mathbb{P}}(\qq),
\ee
as in the perturbative limit of \cite{ACPRA2009}
[that is for the eigenfrequencies close to $\omega_0$ when  $\omega^2_p/\omega_0^2\to 0$, $\omega_p$ being
the plasma frequency]. The resulting density of states is
\begin{equation}
\rho(\omega) = \sum_{n} \int_{\mathcal{D}} \frac{d^3q}{(2\pi)^3} \, 
\delta(\omega-\omega_{\qq,n})
\label{eq:rho}
\end{equation}
where the integral over $\qq$ is taken in the unit cell $\mathcal{D}=\{\sum_{i=1}^3Q_i\tilde{\mathbf{e}}_i, -\frac{1}{2}\leq Q_i<
\frac{1}{2}\}$ of the reciprocal lattice of basis $(\tilde{\mathbf{e}}_i)_{1\leq i\leq 3}$, the sum over $n$ runs over the all the eigenvectors of
$\mathbb{P}(\qq)$  and $\omega_{\qq,n}$ is the corresponding eigenfrequency.

For the Gaussian smoothing function, the limit of the band structure for $b\to 0$ is computed in practice
from the relation 
\be
\label{eq:caspart}
\bar{\mathbb{P}}_{\alpha\mu,\beta\nu}(\qq)
\simeq 
\mathbb{P}_{\alpha\mu,\beta\nu}(\qq) e^{-k_0^2 b^2}
\ee
which holds within an exponentially small error in $(d_{\rm min}/b)^2 \gg 1$ where
$d_{\rm min}$ is the minimal interatomic distance \cite{ACPRL2009,ACPRA2009}. 
Note that this relation, obtained for the particular case of a periodic system, is consistent with the general
result (\ref{eq:relgeni}), and implies that the eigenvectors of $\bar{\mathbb{P}}(\qq)$ essentially coincide with the ones of
$\mathbb{P}(\qq)$.
For the diamond, $d_{\rm min}=a\sqrt{3}/4$. We used typically  $b=0.05a$, to which we applied the above
$b\to 0$ extrapolation formula to
obtain the histogram in Figs.~\ref{fig:dos},\ref{fig:dosvac},\ref{fig:dosvaccube}.

\subsection{Penetration depth for the infinite periodic system \label{ss:pen}}

In this subsection we wish to derive, for an infinite system, the value of the penetration depth $\xi$ and to confirm
that it depends on the considered direction of the direct space and that it diverges at the band edges,
both properties having already been observed for a finite-size system in section \ref{ss:pd}.

Hence, we have to solve Eq.~(\ref{eq:steady}) in presence of a forcing source dipole $d_\alpha^{\mathrm s}=\check{d}_\alpha^{\mathrm s}\,e^{-i\omega_s t}$ placed  in $\rr_s$.
The solutions we look for are the steady state dipole amplitudes $\check{d}_{i,\alpha}=
\check{d}^{(\mu)}_{\RR,\alpha}$ on each diamond lattice site of position $\rr_i=\RR+\rr^{(\mu)}$, where $\RR$ belongs
to the Bravais direct lattice. Since the scope is to determine the penetration length $\xi$, we restrict
ourselves to the case where the source frequency $\omega_s$ is in the band gap.
Then the dipole amplitudes are expected to decay exponentially at large distances, and one may introduce the Fourier
transform
\be
\label{eq:defFour}
\check{d}_{\qq,\alpha}^{(\mu)} = \sum_{\RR\in\mathrm{L}} \check{d}_{\RR,\alpha}^{(\mu)} e^{-i\qq\cdot\RR}.
\ee
One applies this Fourier transform to the spatially smoothed version of Eq.~(\ref{eq:steady}); for a Gaussian smoothing function, the source frequency
is actually chosen to be $\bar{\omega}_s$ given by Eq.~(\ref{eq:relsour}), which ensures that the forced dipole amplitudes
are essentially unaffected by the smoothing. In what follows, we can thus omit the bar (indicating the spatial smoothing) over the dipoles and the penetration
depth.
After calculations that closely resembles the ones of section \ref{ss:dosps}:
\begin{multline}
-\hbar(\bar{\omega}_s-\omega_0) \check{d}_{\qq,\alpha}^{(\mu)} + \sum_{\beta,\nu}
\bar{\mathbb{P}}_{\alpha\mu,\beta\nu}(\qq) \check{d}_{\qq,\beta}^{(\nu)} = \\
-\frac{1}{\mathcal{V}_\mathrm{L}} \sum_{\KK\in\mathrm{RL}} e^{i(\KK+\qq)\cdot(\rr^{(\mu)}-\rr_s)}
\sum_\beta \tilde{\bar{g}}_{\alpha\beta}(\KK+\qq) \check{d}_\beta^s.
\end{multline}
One writes the formal solution of this linear system in terms of the inverse of the matrix 
$\bar{\mathbb{P}}(\qq)-\hbar(\bar{\omega}_s-\omega_0) \openone$, where $\openone$ is the identity; this inverse
exists for all $\qq$ since $\bar{\omega}_s$ is in the band gap of the spatially smoothed model. Then applying the inverse Fourier transform 
\be
d_{\RR,\alpha}^{(\mu)}= \int_\mathcal{D} \frac{d^3q}{\mathcal{V}_{\mathrm{RL}}} d_{\qq,\alpha}^{(\mu)} e^{i\qq\cdot\RR},
\ee
and using $\KK\cdot\RR=0$ modulo $2\pi$, one obtains the forced dipole amplitude on each lattice site:
\begin{multline}
\label{eq:dip_force}
\check{d}^{(\mu)}_{\RR,\alpha}=-\sum_{\beta,\gamma,\nu}\sum_{\KK\in\mathrm{RL}}\int_{\mathcal{D}}\frac{d^3q}{(2\pi)^3} 
\, e^{i(\KK+\qq)\cdot(\RR+\rr^{(\nu)}-\rr_s)} \\
 \left\{\left[\bar{\mathbb{P}}(\qq)-\hbar(\bar{\omega}_s-\omega_0) \openone\right]^{-1}\right\}_{\alpha\mu,\beta\nu}\,\tilde{\bar{g}}_{\beta\gamma}(\KK+\qq)\,\check{d}_{\gamma}^{s}.
 \end{multline}

A first application of Eq.~(\ref{eq:dip_force}) is to evaluate the dipole amplitudes from a numerical integration
over $\qq$ and, fitting them in a region of large values of $R$ in some direction $\uu$, to extract the penetration depth
in that direction. Using up to $256^3$ points in the numerical integration over $\qq$, this leads to the
stars in Fig.~\ref{fig:pendepzoom}, that compare well to the penetration depth extracted from the simulations
on a finite size system in section \ref{ss:pd}.
Furthermore this approach is numerically more efficient close to the borders of the band gap, where the penetration depth
diverges and the finite size effects of the simulations become stronger.

A second strategy to obtain the penetration depth from Eq.~(\ref{eq:dip_force}) is to use the residue theorem.
Since $\KK+\qq$ spans $\mathbb{R}^3$ when $\KK$ spans the reciprocal lattice and $\qq$ spans its unit cell $\mathcal{D}$,
and since $\bar{\mathbb{P}}(\qq)=\bar{\mathbb{P}}(\qq+\KK)$, Eq.~(\ref{eq:dip_force}) can be rewritten as
\begin{multline}
\label{eq:dip_force2}
\check{d}^{(\mu)}_{\RR,\alpha}=-\sum_{\beta,\gamma,\nu} \int_{\mathbb{R}^3} \frac{d^3k}{(2\pi)^3}
\, e^{i\kk\cdot(\RR+\rr^{(\nu)}-\rr_s)} \\
\left\{\left[\bar{\mathbb{P}}(\kk)-\hbar(\bar{\omega}_s-\omega_0) \openone\right]^{-1}\right\}_{\alpha\mu,\beta\nu}\,
\tilde{\bar{g}}_{\beta\gamma}(\kk)\,\check{d}_{\gamma}^{s}.
\end{multline}
To take the large $\RR$ limit in the direction $\uu$, we set 
\be
\RR=r\uu +O(1)\ \ \ \mbox{with}\ r>0.
\ee
We split the integration over $\kk$
into an integral over the component $k_\parallel$ of $\kk$ along $\uu$ and over the transverse components $\kk_\perp$ of
$\kk$. Then $\kk_\perp\cdot \RR$ remains bounded, whereas $\uu\cdot\RR$ is divergent.

First, we consider the integral over $k_\parallel \in \mathbb{R}$ for a fixed $\kk_\perp$. The integrand involves
the exponential factor $e^{ik_\parallel r}$; since $r>0$ we close the integration contour with a half-circle (of diverging radius)
in the upper complex plane
\footnote{To this end, the Gaussian smoothing function $\chi$ is not appropriate. One can rather
take $\chi(\rr)\propto e^{-r/b}/r$, whose Fourier transform is a Lorentzian.}. Whereas the equation for $k_\parallel$:
\be
\label{eq:trans}
\bar{\omega}_{k_\parallel\uu+\kk_\perp,n}=\bar{\omega}_s,
\ee
where $\bar{\omega}_{\kk,n}$ is the dispersion relation of the $n$-th band of eigenfrequencies for the spatially smoothed periodic system,  
has for sure no real solution since $\bar{\omega}_s$ is in the band gap, it may have complex solutions 
$k_\parallel^{(0)}$ with a positive imaginary part.
Due to the occurrence of the inverse matrix involving $\bar{\mathbb{P}}(\kk)$ in the integrand, such complex solutions
provide poles in the half upper plane, which according to the residue theorem lead to the damped exponential
$\exp(i k^{(0)}_{\parallel} r)$. If (\ref{eq:trans}) admits several roots, or roots for various band
index $n$, one has to keep the value $n_0$ of $n$ and the root $k^{(0)}_{\parallel}$ leading to the smallest
imaginary part, that provides the leading contribution in the large $r$ limit.

Then one has to remember that there is still an integral over $\kk_\perp$, and that $k^{(0)}_{\parallel}$ depends
on $\kk_\perp$. We thus face an integral of the form
\be
\label{eq:integrale}
\check{d}^{(\mu)}_{\RR,\alpha}=\int_{\mathbb{R}^2} \frac{d^2k_\perp}{(2\pi)^2} e^{ik_\parallel^{(0)}(\kk_\perp)r}
\frac{f(\kk_\perp)}{\partial_{k_\parallel}\bar{\omega}_{k^{(0)}_\parallel(\kk_\perp)\uu+\kk_\perp,n_0}}
\ee
where the derivative of the band dispersion relation in the denominator originates from the residue
of the pole in $k_\parallel^{(0)}(\kk_\perp)$ and the $r$-independent function $f$ in the numerator is easily
reconstructed from Eq.~(\ref{eq:dip_force2}). To obtain an asymptotic equivalent of the integral (\ref{eq:integrale})
in the large-$r$ limit, we use the saddle-point method: Eq.~(\ref{eq:integrale}) is dominated by
the contribution of the vicinity of the stationary point of the ``phase", that is $\kk_\perp^{(0)}$ such that
\footnote{If there are several stationary points, one has to keep the one leading to the smallest imaginary part of
$k_\parallel^{(0)}$.}
\be
\label{eq:station}
\partial_{\kk_\perp} k_\parallel^{(0)}(\kk_\perp^{(0)}) = \mathbf{0}.
\ee
As we shall see, in general $\kk_\perp^{(0)}$ has complex coordinates (in the plane orthogonal to $\uu$)
and one has to deform the integration domain of (\ref{eq:integrale}) to let the integration go through
the stationary point \footnote{This is why the naive minimization of the imaginary part of
$k_\parallel^{(0)}(\kk_\perp)$ over real-component $\kk_\perp$ only gives an upper bound on the penetration
length in the direction $\uu$.}.
Then one quadratizes the variation of the pole around the stationary point:
\be
k_\parallel^{(0)}(\kk_\perp^{(0)}+\delta\kk_\perp) = k_\parallel^{(0)}(\kk_\perp^{(0)}) + \delta\kk_\perp\cdot B 
\delta\kk_\perp+O(\delta k_\perp^3),
\label{eq:defB}
\ee
where the relevant deviations of $\kk_\perp$ from the stationary point scale as $1/r^{1/2}$.
One finally gets the equivalent
\be
\label{eq:Gauss}
\check{d}^{(\mu)}_{\RR,\alpha} \underset{r\to\infty}{\sim} 
\frac{e^{ik_\parallel^{(0)}(\kk_\perp^{(0)})r} f(\kk_\perp^{(0)})}
{\partial_{k_\parallel}\bar{\omega}_{k^{(0)}_\parallel(\kk_\perp^{(0)})\uu+\kk_\perp^{(0)},n_0}}
\int_{\mathbb{R}^2} \frac{d^2\delta k_\perp}{(2\pi)^2} e^{ir \delta\kk_\perp \cdot B \delta\kk_\perp}
\ee
where the Gaussian integral provides a factor $1/r$. The inverse of the penetration depth 
in direction $\uu$ is thus
\be
\label{eq:kappa_analy}
\kappa(\uu)\equiv \frac{1}{\xi(\uu)} = \mbox{Im}\, \left[k^{(0)}_\parallel(\kk_\perp^{(0)})\right].
\ee

In general, this procedure is however difficult to use, even numerically, as one has to look for poles
of the dispersion relation for a wavevector $\kk^{(0)}$
with three complex coordinates. An important and manageable limiting
case is for a source frequency $\bar{\omega}_s$ very close to the lower border $\bar{\omega}_{\rm inf}$ or the upper border
$\bar{\omega}_{\rm sup}$ of the band gap. The penetration depth is then expected to diverge, so that the imaginary components
of the wavevector are small and its real components are close to the location $\qq_0$ in the Bloch vector space
of the band gap border (such that $\bar{\omega}_{\qq_0,n_0}$ is equal to $\bar{\omega}_{\rm inf}$ or $\bar{\omega}_{\rm sup}$).
One can then quadratize the dispersion relation around the location of the border:
\be
\label{eq:quadra}
\bar{\omega}_{\qq_0+\delta\qq,n_0} =\bar{\omega}_{\qq_0} + \delta\qq \cdot \bar{A} \delta \qq+ O(\delta q^3)
\ee
where $\bar{A}$ (resp.\ $-\bar{A}$) is a positive definite matrix for the upper (resp.\ lower) border of the band gap.
Note that, according to Eq.~(\ref{eq:caspart}),  $\bar{A}$ is related to its zero-$b$ limit $A$, that is to the matrix $A$ of
the original model, by
\be
\bar{A} \simeq  e^{-k_0^2 b^2} A
\ee
within an exponentially small error in $1/b^2$.
Then the solution of (\ref{eq:trans}) obeying the stationarity condition (\ref{eq:station}) can be obtained analytically:
\be
\label{eq:k0}
\kk^{(0)}\equiv k_\parallel^{(0)}(\kk_\perp^{(0)})\uu+\kk_\perp^{(0)} \simeq \qq_0 + i\kappa(\uu)
\frac{\bar{A}^{-1} \uu}{\uu\cdot \bar{A}^{-1} \uu}
\ee
with the expression for the inverse penetration depth
\be
\label{eq:kappa}
\kappa(\uu) = \left[(\bar{\omega}_{\qq_0}-\bar{\omega}_s) \uu\cdot \bar{A}^{-1} \uu \right]^{1/2}
\ee
where $\bar{A}^{-1}$ is the inverse of the matrix $\bar{A}$. In practice, one may find that the band gap border is obtained
for several values of $\qq_0$, due to symmetry properties (as it shall be the case for the diamond lattice).
At fixed direction $\uu$, one then has to select the value of $\qq_0$ leading to the minimal value of
$\kappa(\uu)$ in Eq.~(\ref{eq:kappa}). Eqs.~(\ref{eq:k0},\ref{eq:kappa}) are derived in the Appendix \ref{app:ldp},
where the complete resulting expression for $\check{d}^{(\mu)}_{\RR,\alpha}$ is also given.

A simple consequence of (\ref{eq:kappa}) is the asymptotic expression for the {\sl maximal}
penetration depth at a given frequency $\bar{\omega}_s$, {\sl i.e.\ } maximised over the direction $\uu$, close to
a band gap border:
\be
\xi_{\rm max} \underset{\bar{\omega}_s \to \bar{\omega}_{\rm bord}}{\sim }
\left(\frac{\bar{A}_{\rm max}}{\bar{\omega}_{\rm bord}-\bar{\omega}_s}\right)^{1/2},
\ee
where $\bar{A}_{\rm max}$ is the eigenvalue of the matrix $\bar{A}$ of maximal modulus.

We have explicitly evaluated the prediction (\ref{eq:kappa}) in the vicinity of the
upper border of the band gap. Irrespective of the value of $k_0a$, 
we find that the frequency $\bar{\omega}_{\rm sup}$ of this upper border 
is reached on the so-called $L$ point of the first Brillouin zone of the lattice, corresponding to $\qq_0=(\tilde{\eee}_1+
\tilde{\eee}_2+\tilde{\eee}_3)/2=(\pi/a)(\eee_x+\eee_y+\eee_z)$ [see Eq.~(\ref{eq:tei}) for the values of the
$\tilde{\eee}_i$], as it was already suspected in \cite{ACPRA2009}.
This point is so symmetric that all the six components of the corresponding eigenvector of the matrix $\bar{\mathbb{P}}$
are equal, which leads to the quite explicit expression
\begin{multline}
\bar{\omega}_{\rm sup}-\omega_0=  \frac{\Gamma}{2}
\left[\frac{1+2(k_0b)^2}{2\pi^{1/2} (k_0b)^3}-\textrm{Erfi}\,(k_0b) e^{-k_0^2b^2}\right] \\
+ \frac{2\pi\Gamma}{k_0^3 \mathcal{V}_{\mathrm{L}}}  \sum_{\KK\in \mathrm{RL}} \sum_\nu \cos[\KK'\cdot (\rr^{(\mu)}-\rr^{(\nu)})]
\frac{K'^{2}e^{-K'^{2} b^2}}{k_0^2-K'^{2}} 
\end{multline}
where $\KK'=\KK+\qq_0$.  However, this frequency is also exactly reached for 13 other values of $\qq_0$, so that
\be
\bar{\omega}_{\qq_0}=\bar{\omega}_{\rm sup} \ \mbox{for}\ \ 2\qq_0\in\{\pm \tilde{\eee}_1 \pm \tilde{\eee}_2 \pm \tilde{\eee}_3, \pm 
\tilde{\eee}_1, \pm \tilde{\eee}_2, \pm \tilde{\eee}_3\}.
\ee
For a given direction $\uu$, one thus calculates the $14$ corresponding matrices $\bar{A}$, which are all {\sl similar},
and one keeps the one giving
the smallest contribution to Eq.~(\ref{eq:kappa}). For $\uu=\eee_1$ and $\uu=\eee_x$ this leads to 
the dashed line in the right part of Fig.~\ref{fig:pendepzoom}a and Fig.~\ref{fig:pendepzoom}b respectively,
in excellent agreement with the numerical evaluation of (\ref{eq:dip_force}) and in good agreement
with the finite-size simulations. Furthermore, for $k_0a=2$ as in the simulations,
the direction $\eee_1$ corresponds
to the twice degenerate, maximal modulus eigenvalue $\bar{A}_{\rm max}$ of some of the $14$ matrices $\bar{A}$ 
(the ones associated to $\qq_0=\pm \frac{1}{2} \tilde{\eee}_2$ 
and $\qq_0=\pm \frac{1}{2} \tilde{\eee}_3$) so that the maximal penetration depth $\xi_{\rm max}$ is obtained in that direction
$\eee_1$.
Remarkably, for $k_0 a$ large enough (but smaller than the value $k_0 a\simeq 5.14$ leading to a closure of the gap),
we find that the conclusion changes, and that the maximal penetration depth is now obtained
in the direction $(\eee_x+\eee_y+\eee_z)/\sqrt{3}$.  This change suggests that there exists a magic value of
$k_0a$ such that the matrix $\bar{A}$ is scalar and, close to the upper bord of the band gap, the penetration
depth is isotropic, which is confirmed by the diagonalisation of $\bar{A}$ that leads to
\footnote{For $k_0 a$ below that value, $\bar{A}_{\rm max}$ is twice degenerate and the corresponding eigenspace is the plane 
orthogonal to $(\eee_x+\eee_y +\eee_z)/\sqrt{3}$ for $\qq_0=(\tilde{\eee}_1+ \tilde{\eee}_2+\tilde{\eee}_3)/2$ 
and the plane orthogonal to
$(-\eee_x+\eee_y +\eee_z)/\sqrt{3}$ for $\qq_0=\tilde{\eee}_1/2$. For $k_0 a$ above that value, $\bar{A}_{\rm max}$ is
not degenerate and the corresponding eigenvector is $(\eee_x+\eee_y +\eee_z)/\sqrt{3}$ 
for $\qq_0=(\tilde{\eee}_1+ \tilde{\eee}_2+\tilde{\eee}_3)/2$, and $(-\eee_x+\eee_y +\eee_z)/\sqrt{3}$ for $\qq_0=\tilde{\eee}_1/2$.}:
\be
(k_0 a)_{\rm iso}^{\rm sup}\simeq 2.8632.
\ee

We have also explicitly evaluated the prediction of Eq.~(\ref{eq:kappa}) in the vicinity of the lower border of the band gap.
We have found that the frequency $\bar{\omega}_{\rm inf}$ of this lower border is obtained in 12 values $\qq_0$ of the Bloch vector,
that weakly depend on $k_0 a$ and that can be parameterized in terms of a single positive dimensionless unknown quantity
$\sigma$:
\begin{multline}
\label{eq:q0inf}
\!\! \bar{\omega}_{\qq_0}=\bar{\omega}_{\rm inf} \ \mbox{for}\ \qq_0\!\in\! \{\pm \sigma (\tilde{\eee}_1-\tilde{\eee}_2),
\pm \sigma (\tilde{\eee}_1-\tilde{\eee}_3), \pm \sigma (\tilde{\eee}_2-\tilde{\eee}_3), \\
\pm[\sigma(\tilde{\eee}_1+\tilde{\eee}_2)+(2\sigma-1)\tilde{\eee}_3],
\pm[\sigma(\tilde{\eee}_1+\tilde{\eee}_3)+(2\sigma-1)\tilde{\eee}_2], \\
\pm[\sigma(\tilde{\eee}_2+\tilde{\eee}_3)+(2\sigma-1)\tilde{\eee}_1]
\}
\end{multline}
where the basis vectors of the reciprocal of the fcc lattice are given by Eq.~(\ref{eq:tei}).
Note that the last six elements of (\ref{eq:q0inf}) have a $\sigma$-independent component
$\pm 2\pi/a$ in the Cartesian basis, along $\eee_z$, $\eee_y$, $\eee_x$ respectively, and their components
along the other two Cartesian axes are equal; these six elements are thus located on the straight
line $XU$, where $X$ and $U$ are standard remarkable points of the first Brillouin zone of the diamond lattice.
For the value $k_0a=2$ taken in the figures, we numerically obtained $\sigma\simeq 0.330\, 346$.
For those 12 values of $\qq_0$, we have determined the 12 similar matrices $\bar{A}$ describing the local quadratization
of $\bar{\omega}_\qq$ and we have kept, for a given $\uu$ equal to $\eee_1$ or $\eee_x$, the one giving 
the smallest contribution to Eq.~(\ref{eq:kappa}). This has led to the dashed line in the left part of 
Fig.~\ref{fig:pendepzoom}a and Fig.~\ref{fig:pendepzoom}b respectively, again in excellent agreement with the numerical 
evaluation of (\ref{eq:dip_force}) and in good agreement with the finite-size simulations.
For $k_0a=2$, it is also found that $\eee_1$ is the eigenvector of two of the 12 similar $\bar{A}$ matrices
[the ones corresponding to the last two elements of (\ref{eq:q0inf})] with the non degenerate, largest modulus eigenvalue
$\bar{A}_{\rm max}$, so that the maximal penetration depth $\xi_{\rm max}$ is actually achieved
in that direction, close to the lower border of the band gap.
For larger values of $k_0 a$, the situation can change to a maximal penetration depth obtained along
direction $\eee_x$. This change occurs for the magic value
\be
(k_0 a)^{\rm inf}_{\rm change}\simeq 2.9412
\ee
where $\sigma\simeq 0.353\, 740$ and the maximal modulus eigenvalue $\bar{A}_{\rm max}$ of the matrices $\bar{A}$ is twice
degenerate.

\subsection{States in the gap due to vacancies \label{subsec:vac}}

We now create a single vacancy in the periodic system (still using the spatially smoothed version), by removing the atom
at the location $\rr_{i_0}=\RR_0+\rr^{(\mu_0)}$, that is at the lattice site
$\RR_0$ on the sublattice $\mu_0$. The eigenspectrum of the spatially smoothed version of (\ref{eq:dipoli})
is expected to remain real ($\bar{\gamma}=0$) but there may now be eigenvalues
with $\bar{\omega}$ in the band gap of the periodic system, corresponding to states
exponentially localized around the vacancy. As we will see, the corresponding
$\bar{\omega}$ are given by Eq.~(\ref{eq:cond_dans_gap}).

To look for such in-gap states, we use the following trick:
Starting from a periodic system in presence of a source dipole in $\rr_s$
(of imposed frequency $\bar{\omega}_s$ and amplitudes $\check{d}^s_\alpha$),
we imagine that the vacancy on site $\rr_{i_0}$ results from the coalescence
of the corresponding forced dipole $\check{d}_{i_0,\alpha}$ with the
source dipole in the limit where the source location tends to the location
of the vacancy:
\be
\label{eq:coal}
\lim_{\rr_s\to \rr_{i_0}} \check{d}_{i_0,\alpha} = -\check{d}^s_\alpha,
\ \forall \alpha.
\ee
In this case, the total dipole carried by the vacancy site vanishes, 
as if there was indeed a vacancy there. Obviously, condition
(\ref{eq:coal}) can be satisfied only for specific values of $\bar{\omega}_s$
in the band gap of the spatially smoothed model, that we now determine.

Writing Eq.~(\ref{eq:dip_force}) for $\RR=\RR_0$, $\mu=\mu_0$, 
$\rr_s=\RR_0+\rr^{(\mu_0)}$, and replacing 
$\check{d}_{\RR_0,\alpha}^{(\mu_0)}$ with $-\check{d}^s_\alpha$,
we obtain the homogeneous linear system
\begin{multline}
\label{eq:interm}
\check{d}^s_\alpha =  \sum_{\beta,\gamma,\nu} \int_{\mathcal{D}}
\frac{d^3q}{\mathcal{V}_{\mathrm{RL}}} 
\left\{[\bar{\mathbb{P}}(\qq)-\hbar(\bar{\omega}_s-\omega_0)\openone]\right\}_{\alpha\mu_0,
\beta\nu} \\
\times \bar{\mathbb{Q}}_{\beta\nu,\gamma\mu_0}(\qq) \check{d}_\gamma^{s} 
\end{multline}
where we used $(2\pi)^3=\mathcal{V}_{\mathrm{RL}} \mathcal{V}_{\mathrm{L}}$
and we called $\bar{\mathbb{Q}}_{\alpha\mu,\beta\nu}(\qq)$ 
the non-scalar contribution to $\bar{\mathbb{P}}_{\alpha\mu,\beta\nu}(\qq)$,
that is the second contribution in the right-hand side of
Eq.~(\ref{eq:cas_pertu}). In terms of matrices, 
\be
\label{eq:decPQ}
\bar{\mathbb{P}}(\qq)= \Lambda \openone + \bar{\mathbb{Q}}(\qq),
\ee
where $\Lambda$ is the coefficient of the scalar contribution, that is of the first term in Eq.~(\ref{eq:cas_pertu}),
$\Lambda=-[\bar{g}_{\alpha\alpha}(\mathbf{0})+i\hbar\bar{\Gamma}/2]$
(this is independent of the direction $\alpha$).
We recognize a matrix product in Eq.~(\ref{eq:interm}), related to the sum
over $\nu$ and $\beta$; we then use
\be
[\bar{\mathbb{P}}(\qq)-\hbar\bar{\delta} \openone]^{-1} \bar{\mathbb{Q}}(\qq) = \openone
+(\hbar\bar{\delta}-\Lambda)[\bar{\mathbb{P}}(\qq)-\hbar\bar{\delta} \openone]^{-1},
\ee
with $\bar{\delta}=\bar{\omega}_s-\omega_0$.
The contribution to (\ref{eq:interm}) of $\openone$ in that expression
exactly reproduces the term $\check{d}^s_\alpha$ of the left-hand side
of (\ref{eq:interm}), since the integral over $\qq$ on the primitive cell
$\mathcal{D}$ of the reciprocal lattice is equal to $\mathcal{V}_{\mathrm{RL}}$.
Simplifying the remaining contribution by the factor $\hbar\bar{\delta}-\Lambda$, it remains
\be
\label{eq:cond_dans_gap}
0 = \int_{\mathcal{D}} \frac{d^3q}{\mathcal{V}_{\mathrm{RL}}}
\sum_{\gamma} 
\left\{[\bar{\mathbb{P}}(\qq)-\hbar(\bar{\omega}_s-\omega_0)\openone]^{-1}\right\}_{\alpha\mu_0,
\gamma \mu_0}  \check{d}_\gamma^{s},  \ \forall \alpha.
\ee
This must have a non-zero solution for the source dipole, which is equivalent
to requiring that the $\bar{\omega}_s$-dependent $3\times 3$ hermitian matrix
in (\ref{eq:cond_dans_gap}) has a zero eigenvalue.
To show that the condition (\ref{eq:cond_dans_gap}) is not only sufficient,
but also necessary, we have performed an alternative calculation, presented
in Appendix~\ref{app:sic}, that has also the advantage of including
the case of several vacancies.

For the diamond lattice, we have evaluated numerically the integral
over the Bloch vector $\qq$  in Eq.~(\ref{eq:cond_dans_gap}).
We then find that the resulting $3\times 3$ hermitian matrix is scalar.
As the eigenvalues of that matrix are increasing functions of $\bar{\omega}_s$, 
as can be shown with the Hellmann-Feynman theorem, this implies
that there is at most one solution for $\bar{\omega}_s$ in the band gap.
Numerically, we find that there is a solution, whose value [after extrapolation to $b\to 0$
using Eq.~(\ref{eq:relgeni})] for $k_0a=2$
is indicated by a vertical dotted line in Fig.~\ref{fig:dosvac}, in agreement with
a peak location in the density of states in the numerical simulations.

In the case of several vacancies, we can extend our analysis as described in appendix~\ref{app:sic}.
By numerical solution of Eq.~(\ref{eq:systemelac}), we have then investigated the in-gap states for two vacancies on sites separated
by $\breve{R}_2-\breve{R}_1=\mathbf{0}, \eee_1$ or $a\eee_x$, being either on the same 
sublattice ($\breve{\mu}_1=\breve{\mu}_2$) or on different sublattices  ($\breve{\mu}_1\neq \breve{\mu}_2$).
In most cases, we have found allowed frequencies close to the one of the single-vacancy state, within the width
of the central peak in the inset of Fig.~\ref{fig:dosvac}; those states can not be resolved in that figure and we have not indicated them.
For the two geometries specified in the caption of Fig.~\ref{fig:dosvac}, we have found frequencies of two-vacancy states
that are clearly out of the central peak, see the red and blue vertical dotted lines; in particular, the prediction
with $(\omega-\omega_0)/\Gamma\simeq -4$ seems to match quite well the very clear secondary peak that emerges
in the figure for increasing concentration of vacancies.

\section{Conclusion
\label{sec:conclusion}}

Three-dimensional periodic arrangements of extended scattering objects leading to an omnidirectional
band gap for light have been known since the 90's, starting from the diamond lattice configuration of dielectric microspheres of
 \cite{Soukoulis}. In the case of a periodic ensemble of {\sl point-like} scatterers, the technical issues
affecting the calculation of the band structure of light have been
solved only recently \cite{Knoester06,ACPRA2009,ACPRL2009}, which has allowed to show that the diamond lattice can also lead to a photonic band gap in the
point-like case \cite{ACPRA2009}.

With cold atom experiments, a diamond-like ensemble of point-like scatterers is in principle realizable, provided that
one produces, in the appropriate optical lattice geometry \cite{John04,ACPRA2009}, a high quality Mott phase of atoms 
\cite{Bloch02,Phillips07} having an optical transition
between a spin zero ground state and a spin one electronic excited state \cite{Takahashi09}.
In practical realizations, there will be of course unavoidable deviations from the ideal infinite periodic case, that we have
quantified in the present work with numerical solutions of linearly coupled dipoles equations with about
$3\times 10^4$ particles.

A first issue is due to effects of the finite size of the atomic medium. Rather than having a band structure, light has a continuous spectrum
of scattering states; by analytic continuation to the lower half of the complex plane, however, it is more physical to consider, as we have done,
the discrete complex eigenfrequencies $\omega-i\gamma$ of the {\sl resonances} of the system. In the distribution function of $\omega$,
the forbidden gap remains visible in our simulations. It remains actually quite visible if one restricts to the resonances
with a half decay rate $\gamma$ much smaller than the free space single atom spontaneous emission rate $\Gamma$; 
such a filtering of the resonances could be realized experimentally by performing a frequency measurement after
an adjustable time delay, during which the short-lived resonances decay and are suppressed.
Amusingly, a narrow peak in the distribution function of $\omega$ was observed close to the center of the infinite system band gap, 
when the finite size atomic medium has a cubic shape;
such a peak, absent when the medium has a spherical shape, is a very clear finite size effect.

A second issue is due to vacancies inside the atomic medium. For a concentration of a few per cent of vacancies, narrow peaks
emerge in the distribution function of $\omega$ inside the gap. We were able to identify several of
these peaks as corresponding to the frequencies of localised states around one or two close vacancies in an otherwise infinite periodic medium.
At higher concentrations of vacancies, e.g.\ 20\%, with no filtering on $\gamma$, the gap disappears.

From our finite size sample, we have shown that one can quite accurately extract the penetration depth $\xi$ of the light in the medium,
and that the obtained values compare well with independent calculations in a periodic medium. Away from the borders of the band gap,
$\xi$ as a function of the imposed field frequency $\omega_s$ exhibits a plateau at a remarkably low value, between $0.5 a$ and $a$, 
where $a$ is the lattice constant of the underlying fcc lattice. 
Close to the borders $\omega_{\rm bord}$ of the band gap, one can even directly observe, in our finite
size system, the onset of the divergence of
$\xi$ as $1/|\omega_s-\omega_{\rm bord}|^{1/2}$, with a prefactor close to our analytical predictions. 
We have also observed from the simulations  that $\xi$ is anisotropic (it depends on the direction of space), 
in agreement with our theoretical analysis,
and that this anisotropy becomes quite pronounced close to the lower border of the band gap.

\acknowledgments
We acknowledge a discussion with Dominique Delande at an early stage of this project.
M.A.\ is member of the LabEx NUMEV.

\appendix

\section{Penetration depth} \label{app:ldp}

In this Appendix, for the spatially smoothed model, we derive the results (\ref{eq:k0},\ref{eq:kappa}) for the penetration depth in the direction $\uu$
at a frequency $\bar{\omega}_s$ close to a border of the band gap, which justifies the use of the quadratized dispersion relation
(\ref{eq:quadra}) around the Bloch vector $\qq_0$, and we give the large-distance equivalent of the forced dipole amplitude,
as obtained from the saddle-point method.

As short-hand notations, we introduce $z=k_\parallel-q_{0\parallel}$ and $\xx=\kk_\perp-\qq_{0\perp}$
as the components along $\uu$ and in the plane orthogonal to $\uu$ of the vector $\kk-\qq_0$.
We also introduce the frequency deviation from the nearest band border, $\bar{\Delta}\equiv \bar{\omega}_{\qq_0}-\bar{\omega}_s$.
Then Eq.~(\ref{eq:trans}) reduces to a degree-two equation for $z$:
\be
\label{eq:trinome}
z^2 \uu\cdot \bar{A} \uu + 2 z \xx\cdot \bar{A} \uu + \xx\cdot \bar{A} \xx +\bar{\Delta}=0
\ee
Furthermore, $z$ has to be stationary with respect to a variation of $\kk_\perp$, see Eq.~(\ref{eq:station}). Differentiating
the trinomial (\ref{eq:trinome}) with respect to $\xx$, and using $\partial_\xx z=\mathbf{0}$, one obtains
the vectorial equation
$z Q \bar{A} \uu + Q\bar{A}Q \xx =\mathbf{0}$
where $Q$ projects orthogonally to $\uu$. The solution is 
\be
\xx = - z (Q\bar{A}Q)^{-1} \bar{A}\uu
\ee
where the matrix inverse is intended within the vectorial plane orthogonal to $\uu$. Inserting this solution
into Eq.~(\ref{eq:trinome}) and using $[P\bar{A}P-P\bar{A}Q\, (Q\bar{A}Q)^{-1}\, Q\bar{A}P] P \bar{A}^{-1} P=P$ where $P=1-Q$ is the orthogonal projector
on $\uu$ (see relation (B.23) of \S III.B.2 in \cite{CCT}), one obtains
\be
z=i\kappa(\uu) \ \ \mbox{with} \ \ \kappa(\uu)\ \ \mbox{given by Eq.~(\ref{eq:kappa})}
\ee
Similarly, injecting the closure relation $P+Q=1$, one finds $\bar{A}[\uu-(Q\bar{A}Q)^{-1} \bar{A} \uu]=[P\bar{A}P-P\bar{A}Q\, (Q\bar{A}Q)^{-1}\, Q\bar{A}P] \uu=
(P\bar{A}^{-1}P)^{-1} \uu=\uu/(\uu\cdot \bar{A}^{-1} \uu)$. This gives as in Eq.~(\ref{eq:k0}):
\be
\uu - (Q\bar{A}Q)^{-1} \bar{A} \uu = \frac{\bar{A}^{-1}\uu}{\uu\cdot \bar{A}^{-1} \uu}
\ee

To determine the residue appearing in (\ref{eq:Gauss}), one takes the derivative of the trinomial (\ref{eq:trinome})
with respect to $z$ for a fixed $\xx$. Using the previous relations one obtains
\be
\partial_{k_\parallel}\bar{\omega}_{\kk^{(0)},n_0}=2z \uu\cdot \bar{A}\uu +2\xx\cdot \bar{A}\xx=\frac{2i\kappa(\uu)}{(\uu\cdot \bar{A}^{-1} \uu)}.
\ee
Next, we determine the matrix $B$ in Eq.~(\ref{eq:Gauss}) originating from the quadratization of $z$ around the stationary
point $\xx$. A first order variation $\delta\xx$ induces a second order variation $\delta z$. Performing these variations
in Eq.~(\ref{eq:trinome}) up to second order in $\delta\xx$ and up to first order in $\delta z$,
and using the previous relations, we obtain
\be
B = \frac{i}{2\kappa(\uu)} (\uu\cdot \bar{A}^{-1} \uu) Q \bar{A}Q.
\ee
We conclude that the matrix $i B$ appearing in the Gaussian integral (\ref{eq:Gauss}) is negative, which justifies
the fact that the saddle point is approached along the real axis direction as in (\ref{eq:Gauss}).
If one performs the Gaussian integral, Eq.~(\ref{eq:Gauss}) reduces to
\be
\check{d}^{(\mu)}_{\RR,\alpha} \underset{r\to\infty}{\sim} \frac{e^{-\kappa(\uu)r}f(\kk_\perp^{(0)})}{4i\pi r}
[\det(Q\bar{A}Q)]^{-1/2}.
\ee
The determinant in that expression is conveniently transformed as 
$\det(Q\bar{A}Q) = (\uu\cdot \bar{A}^{-1} \uu) \det \bar{A}$ using the expression of the matrix of $\bar{A}^{-1}$ in terms of the comatrix of $\bar{A}$
(in an orthonormal basis containing the direction $\uu$).

To obtain our final asymptotic form for the forced dipole amplitude, we note that, for any acceptable vector $\kk^{(0)}$ 
of the pole plus saddle-point analysis, $\kk^{(0)}+\KK$ is again acceptable, where $\KK$ is any vector of
the reciprocal lattice; this is due to the periodicity of the dispersion relation $\bar{\omega}_{\kk,n_0}$. We also include
a sum over possibly degenerate Bloch vector $\qq_0$ leading to the same value $\bar{\omega}_{\qq_0,n_0}$
(as discussed in the main text). We also note that, when $R\to +\infty$,
\begin{multline}
||(\bar{A}^{-1}\bar{\Delta})^{1/2}(\RR+\rr^{(\nu)}-\rr_s)|| = -i (\kk^{(0)}-\qq_0)\cdot (\RR+\rr^{(\nu)}-\rr_s) \\+o(1)
\end{multline}
which gives a simple physical interpretation to the expression (\ref{eq:k0}) of $\kk^{(0)}$: The apparently obscure correction
to $\qq_0$ in (\ref{eq:k0}) simply originates from the fact that what more precisely matters in the asymptotic behavior
of the dipole amplitudes is not $r\uu$ but really the vectorial distance $\RR+\rr^{(\nu)}-\rr_s$ between the considered lattice site
and the source. Finally, we obtain,
for $\bar{\omega}_s$ close to a border of the band gap, the asymptotic equivalent for $R\to \infty$:
\begin{multline}
\check{d}^{(\mu)}_{\RR,\alpha}\sim -
\sum_{\qq_0} \left(\frac{\bar{\Delta}}{\det \bar{A}}\right)^{1/2}\!\!\!\! \frac{e^{i\qq_0\cdot\RR} e^{-||(\bar{A}^{-1}\bar{\Delta})^{1/2}(\RR+\rr^{(\nu)}-\rr_s)||}}
{4\pi\hbar||(\bar{A}^{-1}\bar{\Delta})^{1/2}(\RR+\rr^{(\nu)}-\rr_s)||} \\
\times 
\sum_{\beta,\mu,\nu} 
\sum_{\KK\in\mathrm{RL}} e^{i(\qq_0+\KK)\cdot (\rr^{(\nu)}-\rr_s)} 
\phi_{\alpha\mu}^{(n_0)} \phi_{\beta \nu}^{(n_0)*}
\tilde{\bar{g}}_{\beta\gamma} (\qq_0+\KK) \check{d}^{s}_{\gamma},
\label{eq:dlong}
\end{multline}
where $\phi_{\alpha\mu}^{(n_0)}$ are the components of the normalized eigenvector of $\bar{\mathbb{P}}(\qq_0)$ of eigenvalue 
$\bar{\omega}_{\qq_0,n_0}$, we approximated $\kk^{(0)}$ by $\qq_0$ in the argument of $\tilde{\bar{g}}_{\beta\gamma}$, 
and the square root $(\bar{A}^{-1}\bar{\Delta})^{1/2}$ of the matrix $\bar{A}^{-1} \bar{\Delta}$ is well defined
since this matrix is positive.
Note that the second line of (\ref{eq:dlong}) does not depend on $\RR$.

\section{A general vacancy calculation} \label{app:sic}

We consider here the infinite periodic system, with a finite number of vacancies at nodes $(\breve{\RR}_i,\breve{\mu}_i)$,
$1\leq i\leq n$, where we recall that $\RR$ belongs to the fcc Bravais lattice and $\mu$ labels the sublattices.
The scope is to determine the frequencies $\bar{\omega}$ of the localised states that can exist, due to the presence
of the vacancies, in the band gap of the periodic system, in the spatially smoothed version of the model.

The idea is to formally introduce, in the coupled equations for the dipoles, fictitious dipoles carried
by the vacancies. Among the physical dipoles, the spatially smoothed version of Eq.~(\ref{eq:dipoli}) holds:
\be
\label{eq:dphys}
0=(\Lambda-\hbar\delta) d_{\RR}^{(\mu)} + \sum'_{\RR',\mu'} \bar{g}(\RR+\rr_\mu-\RR'-\rr_{\mu'}) d_{\RR'}^{(\mu')}.
\ee
Here the prime over the summation symbol means that the sum is restricted to the physical dipoles,
$\bar{\delta}=\bar{\omega}-\omega_0$ is the detuning from the atomic resonance, $\Lambda$ is defined below Eq.~(\ref{eq:decPQ}),
and we have used for conciseness
an implicit vectorial notation for the dipoles and an implicit  matrix notation for $\bar{g}$.
For the fictitious dipoles, the equation is that they are equal to zero:
\be
0=(\Lambda-\hbar\bar{\delta}) d_{\breve{\RR}_i}^{(\breve{\mu}_i)}, \ \ \forall i\in \{1,\ldots,n\}.
\ee
This allows to formally extend the sum in Eq.~(\ref{eq:dphys}) to the fictitious dipoles, that
is one can remove the prime over the summation symbol.
One can then merge the two series of equations using the usual plus-minus trick:
for {\sl all} $\RR$ in the Bravais lattice and for {\sl all} sublattices $\mu$, one requires that
\begin{multline}
0=(\Lambda-\hbar\bar{\delta}) d_{\RR}^{(\mu)}+\sum_{\RR',\mu'} \bar{g}(\RR+\rr_\mu-\RR'-\rr_{\mu'}) d_{\RR'}^{(\mu')} \\
-\sum_{i=1}^{n} \delta_{\RR,\breve{\RR}_i} \delta_{\mu,\breve{\mu}_i} \bar{s}_i
\label{eq:syn}
\end{multline}
where $\delta$ is the Kronecker symbol and we have introduced the auxiliary unknowns
\be
\label{eq:defauxi}
\bar{s}_i \equiv \sum_{\RR',\mu'} \bar{g}(\breve{\RR}_i+\rr_{\breve{\mu}_i}-\RR'-\rr_{\mu'}) d_{\RR'}^{(\mu')}.
\ee
Then taking the Fourier transform (\ref{eq:defFour}) of Eq.~(\ref{eq:syn}) and using (\ref{eq:Pini}):
\be
0=\sum_{\mu'} [\bar{\mathbb{P}}(\qq)-\hbar\bar{\delta}\openone]_{\mu\mu'} d_{\qq}^{(\mu')} -
\sum_{i=1}^{n} e^{-i\qq\cdot\breve{\RR}_i} \delta_{\mu,\breve{\mu}_i} \bar{s}_i.
\ee
Since the frequency $\bar{\omega}$ is in the gap, the matrix is invertible, and taking the inverse Fourier transform,
one obtains
\be
\label{eq:drmuexpli}
d_{\RR}^{(\mu)} = \sum_{i=1}^{n} \int_{\mathcal{D}} \frac{d^3q}{\mathcal{V}_{RL}} e^{i\qq\cdot(\RR-\breve{\RR}_i)}
\{[\bar{\mathbb{P}}(\qq)-\hbar\bar{\delta}\openone]^{-1}\}_{\mu\breve{\mu}_i} \bar{s}_i.
\ee

Expressing the fact that the fictitious dipoles are all equal to zero, we find the homogeneous
system of equations:
\be
\label{eq:systemelac}
\sum_{i=1}^{n} \int_{\mathcal{D}} \frac{d^3q}{\mathcal{V}_{RL}} e^{i\qq\cdot(\breve{\RR}_j-\breve{\RR}_i)}
\{[\bar{\mathbb{P}}(\qq)-\hbar\bar{\delta}\openone]^{-1}\}_{\breve{\mu}_j\breve{\mu}_i} \bar{s}_i=0,
\ee
to be satisfied $\forall j\in \{1,\ldots,n\}$. The acceptable in-gap frequencies are such that the system
admits a non-identically zero solution $(\bar{s}_i)_{1\leq i\leq n}$, that is the determinant of the corresponding
$3n\times 3n$ matrix must vanish.
In the case of a single vacancy, this reproduces Eq.~(\ref{eq:cond_dans_gap}).

Finally we have performed the consistency check that, if one replaces in Eq.~(\ref{eq:defauxi})
the dipoles in terms of the auxiliary unknowns $\bar{s}_j$, as given by (\ref{eq:drmuexpli}), one recovers
exactly the same system as (\ref{eq:systemelac}), using Eqs.~(\ref{eq:Pini},\ref{eq:decPQ})
and the fact that the integral over $\qq$ on the primitive cell $\mathcal{D}$ of the reciprocal lattice
is equal to its volume $\mathcal{V}_{RL}$.


\begin{thebibliography}{0}
\expandafter\ifx\csname natexlab\endcsname\relax\def\natexlab#1{#1}\fi
\expandafter\ifx\csname bibnamefont\endcsname\relax
  \def\bibnamefont#1{#1}\fi
\expandafter\ifx\csname bibfnamefont\endcsname\relax
  \def\bibfnamefont#1{#1}\fi
\expandafter\ifx\csname citenamefont\endcsname\relax
  \def\citenamefont#1{#1}\fi
\expandafter\ifx\csname url\endcsname\relax
  \def\url#1{\texttt{#1}}\fi
\expandafter\ifx\csname urlprefix\endcsname\relax\def\urlprefix{URL }\fi
\providecommand{\bibinfo}[2]{#2}
\providecommand{\eprint}[2][]{\url{#2}}

\end{thebibliography}


\begin{thebibliography}{90}

\bibitem{Pastori}
G. Grosso, G. Pastori-Parravicini, \emph{Solid State Physics} 
(Academic Press, 2000).

\bibitem{Kagan} 
A.M. Afanas'ev and Yu. Kagan, Sov. Phys. JETP {\bf 25}, 124 (1967);
G.B. Smirnov, Y.V. Shvydko, JETP Letters {\bf 35}, 505 (1982).

\bibitem{Hopfield58} 
J.J. Hopfield, Phys. Rev. {\bf 112}, 1555 (1958);
V. Agranovich, Sov. Phys. JETP {\bf 37}, 307 (1960).

\bibitem{bookDidier} 
J.D. Joannopoulos, S.G. Johnson, J.N. Winn, and R.D. Meade, \emph{	
Photonic Crystals:
Molding the Flow of Light} [Princeton University Press, Princeton, NJ, 2008] (2nd Edition);
F. Zolla, G. Renversez, A. Nicolet, B. Kuhlmey, S. Guenneau, D. Felbacq, A. Argyros, and S. Leon-Saval, \emph{Foundations of Photonic Crystal Fibres} [Imperial College Press, London, 2012]  (2nd Edition).

\bibitem{Bloch02} 
M. Greiner, O. Mandel, T. Esslinger, T.W. H\"{a}nsch, and I. Bloch, 
Nature {\bf 415}, 39 (2002).

\bibitem{Phillips07}
M. Anderlini, P.J. Lee, B.L. Brown, J. Sebby-Strabley,
W.D. Phillips, J.V. Porto, Nature {\bf 448}, 452 (2007).

\bibitem{ACPRL2009} 
M. Antezza and Y. Castin, \prl {\bf 103}, 123903 (2009).

\bibitem{ACPRA2009} 
M. Antezza and Y. Castin, \pra {\bf 80}, 013816 (2009).

\bibitem{ultraclock1}
Masao Takamoto, Feng-Lei Hong, Ryoichi Higashi, Hidetoshi Katori, Nature {\bf 435}, 321 (2005).

\bibitem{ultraclock2}
T.L. Nicholson, M.J. Martin, J.R. Williams, B.J. Bloom, M. Bishof, M.D. Swallows, S.L. Campbell, and J. Ye,  
Phys. Rev. Lett. {\bf 109}, 230801 (2012) , and references therein.

\bibitem{ultraclock3}
A.D. Ludlow, T. Zelevinsky, G.K. Campbell, S. Blatt, M.M. Boyd, M.H.G. de Miranda, M.J. Martin, J.W. Thomsen, S.M. Foreman, Jun Ye, T.M. Fortier, J.E. Stalnaker, S.A. Diddams, Y. Le Coq, Z. W. Barber, N. Poli, N.D. Lemke, K.M. Beck, and C.W. Oates,  Science {\bf 139}, 1805 (2008).

\bibitem{Coevorden96} 
D.V. van Coevorden, R. Sprik, A. Tip, and A. Lagendijk, Phys. Rev. Lett.  {\bf 77}, 2412 
(1996).

\bibitem{LagendijkRMP}
P. de Vries, D.V. van Coevorden, A. Lagendijk, Rev. Mod. Phys. {\bf 70}, 447 (1998).


\bibitem{Knoester06} 
J.A. Klugkist, M. Mostovoy, and J. Knoester, Phys. Rev. Lett. {\bf 96}, 163903 (2006).

\bibitem{1DGuerin} A. Schilke, C. Zimmermann, P. W. Courteille, and W. Guerin 
\prl {\bf 106}, 223903 (2011); A. Schilke, C. Zimmermann, and W. Guerin, \pra {\bf 86}, 023809 (2012). 

\bibitem{1DGuerinlaser}A. Schilke, C. Zimmermann, P. W. Courteille, and W. Guerin, Nature Photonics {\bf 6}, 101 (2011)

\bibitem{Morigi10} S. Rist, C. Menotti, and G. Morigi, \pra {\bf 81}, 013404 (2010).

\bibitem{Carusotto08}I. Carusotto,
M. Antezza, F. Bariani, S. De Liberato, and C. Ciuti
Phys. Rev. A {\bf 77}, 063621 (2008).

\bibitem{Ritsch07} H. Zoubi, H. Ritsch, \pra {\bf 76}, 013817 (2007).

\bibitem{Morice95}
O. Morice, Y. Castin and J. Dalibard, Phys. Rev. A {\bf 51}, 3896 (1995).

\bibitem{SPIE} 
D. Felbacq, and M. Antezza,  SPIE Newsroom (2012) [DOI: 10.1117/2.1201206.004296], and references therein.

\bibitem{Fano56} 
U. Fano, Phys. Rev. {\bf 103}, 1202 (1956).

\bibitem{Courtois99}
A. Chelnokov, S. Rowson, J.-M. Lourtioz, V. Berger,
J.-Y. Courtois, J. Opt. A: Pure Appl. Opt. {\bf 1}, L3 (1999).

\bibitem{John04}  
O. Toader, T.Y. Chan, and S. John, Phys. Rev. Lett. {\bf 92}, 043905 (2004).

\bibitem{YUPRA2011}
D. Yu, \pra {\bf 84}, 043833 (2011).

\bibitem{scalar}{It is worth stressing that the often used scalar model for light has the evident advantage of drastically reducing the numerical effort, but also the disadvantage of providing a qualitatively and quantitatively wrong description of the physical system. For instance, it is possible to show that already for a simple cubic atomic lattice, the scalar model, in contradiction with the vectorial one, predicts the presence of a band gap.}

\bibitem{Wilk}
Y. Bidel, B. Klappauf, J.C. Bernard, D. Delande, G. Labeyrie, C. Miniatura, D. Wilkowski and R. Kaiser,
Phys. Rev. Lett. {\bf 88}, 203902 (2002).

\bibitem{Jackson}
J.D. Jackson, {\sl Classical Electrodynamics}, 2nd ed.\ (Wiley, New York, 1975).

\bibitem{CCT} C. Cohen-Tannoudji, J. Dupont-Roc, G. Grynberg,
{\sl Processus d'interaction entre photons et atomes}, InterEditions/Editions
du CNRS (Paris, 1988).

\bibitem{Soukoulis}
K.M. Ho, C.T. Chan, C.M. Soukoulis, Phys. Rev. Lett. {\bf 65}, 3152 (1990).

\bibitem{Takahashi09}  
T. Fukuhara, S. Sugawa, M. Sugimoto, S. Taie, Y. Takahashi, 
Phys. Rev. A {\bf 79}, 041604 (2009).

\end{thebibliography}
\end{document}